\documentclass[aps,prc,twocolumn,floatfix]{revtex4}
\usepackage{epsfig}
\usepackage{amsmath}

\begin{document}

\title{Parton shadowing and $J/\psi$-to-Drell-Yan ratio in nuclear collisions at SPS and FAIR}

\author{Partha Pratim Bhaduri}
\affiliation{Variable Energy Cyclotron 
Centre, 1/AF Bidhan Nagar, Kolkata 700 064, India}  
\author{A. K. Chaudhuri}
\affiliation{Variable Energy Cyclotron 
Centre, 1/AF Bidhan Nagar, Kolkata 700 064, India} 
\author{Subhasis Chattopadhyay}
\affiliation{Variable Energy Cyclotron 
Centre, 1/AF Bidhan Nagar, Kolkata 700 064, India}            
\date{\today}
\begin{abstract}

We have analyzed the data on $J/\psi$-to-Drell-Yan production cross section ratio in proton-nucleus ($p+A$) and nucleus-nucleus ($A+A$) collisions, measured by the NA50 collaboration in the SPS energy domain. Two component QVZ model has been employed to calculate $J/\psi$ production cross sections. For both $J/\psi$ and Drell-Yan production, nuclear modifications to the free nucleon structure functions are taken into account. In heavy-ion collisions, such modifications are assumed to be proportional to the local nuclear density resulting in centrality dependent initial state effects. Differences in quark and gluon shadowing leads to a new source of impact parameter dependence of the $J/\psi$ to Drell-Yan production ratio. For $J/\psi$, final state interaction of the produced $c\bar{c}$ pairs with the nuclear medium is also taken into account. A satisfactory description of the data in both $p+A$ and $A+A$ collisions is obtained. Role of the shadowing corrections are investigated in detail. Model calculations are extrapolated to predict the centrality dependence of $J/\psi$-to-Drell-Yan ratio in the FAIR energy regime.

\end{abstract}

%\pacs{25.75.-q,12.38.Mh}
\maketitle
\section{Introduction}

The primary objective of the relativistic heavy-ion collision experiments is to study the nuclear matter under extreme conditions of temperature and density in the laboratory. Of particular interest is the QCD predicted transition from the color confined hadronic phase to color de-confined quark-gluon plasma (QGP) phase. However the transient nature of the plasma renders its identification very difficult. It necessitates the search for experimental evidences that would survive the complex evolution through the later stages of the collision. $J/\psi$ suppression in nuclear collisions had long been identified as a diagnostic tool to indicate the occurrence of the deconfinement transition~\cite{MS,Satz,Vogt}. However subsequent experimental measurements found a considerable reduction of the $J/\psi$ yield already present in proton-nucleus ($p+A$) collisions, which is attributed to the cold nuclear medium of the target nucleus. It is commonly expected that in addition to the cold nuclear matter (CNM) suppression, in nuclear collisions, formation of QGP would lead to anomalous $J/\psi$ suppression. At SPS, $J/\psi$ suppression in heavy-ion collisions is measured at a beam energy 158 A GeV, in Pb+Pb collisions by NA50 collaboration~\cite{Ale05} and in In+In collisions by NA60 collaboration~\cite{Arn07}, in the same kinematic domain. Originally NA50 published their data on $J/\psi$ suppression in terms $J/\psi$-to-Drell-Yan (DY) ratio at different collision centralities. Due to huge statistical uncertainties in measurement of Drell-Yan di-muons, data from NA60 collaboration were published in terms of the measured-by-expected ratio of $J/\psi$ cross sections as a function collision centrality, where expected denotes the $J/\psi$ production cross section evaluated incorporating the nuclear absorption scenario following Glauber formalism of nuclear scattering theory~\cite{LABref}. One obvious disadvantage in this case is that the experimental data become sensitive to the theoretical model inputs. Recently both In+In and Pb+Pb data are presented in terms of the nuclear modification factor ($R_{AA}$)~\cite{Arn11} defined as the ratio of $J/\psi$ production cross sections in $A+A$ ($\sigma_{AA}^{J/\psi}$) and $p+p$ ($\sigma_{pp}^{J/\psi}$) collisions. In this case as well, the $p+p$ measurements were not performed; rather $\sigma_{pp}^{J/\psi}$ was obtained through extrapolation of the parametrized 158 GeV $p+A$ data up to $A=1$. In our previous work~\cite{partha2}, we have shown that the both the data sets on centrality dependence of $R_{AA}^{J/\psi}$, in In+In and Pb+Pb collisions can be reasonably described by the adapted version of the QVZ approach~\cite{Qui}. In the present article, we aim to check the viability of our model by examining the available data corpus on $J/\psi$-to-Drell-Yan ratio in $p+A$~\cite{NA50-400, NA50-450} and Pb+Pb~\cite{Ale05} collisions measured by NA50 collaboration, at SPS energies. The advantage of analyzing this ratio is that the cross-sections for both the processes are directly measured in the experiment. Moreover since the muon pairs for both events were collected in the same run, it certainly gives a better control over the systematic uncertainties. A special emphasis is put on the role of parton shadowing in deciding the behavior of $J/\psi$-to-Drell-Yan ratio observed in nucleon-nucleus and nucleus-nucleus collisions, and its sensitivity to the differences in quark and gluon shadowing parameters. Such effects of parton shadowing on $J/\psi$ production in $p+A$ and $A+A$ collisions have also been investigated earlier~\cite{shadow1, Eme99, Kle03, shadow2}, within the energy domains of SPS, RHIC and LHC. Shadowing factors have been evaluated as a function of rapidity and collision centrality. Our analysis is motivated by and akin to such previous studies as far as evaluation of shadowing effects are concerned. In addition we have also taken into account final state nuclear dissociation of $J/\psi$ in order to explain the measured data. 

\begin{figure*} 
\scalebox{0.25}
{
\includegraphics[width=1.3 \textwidth]{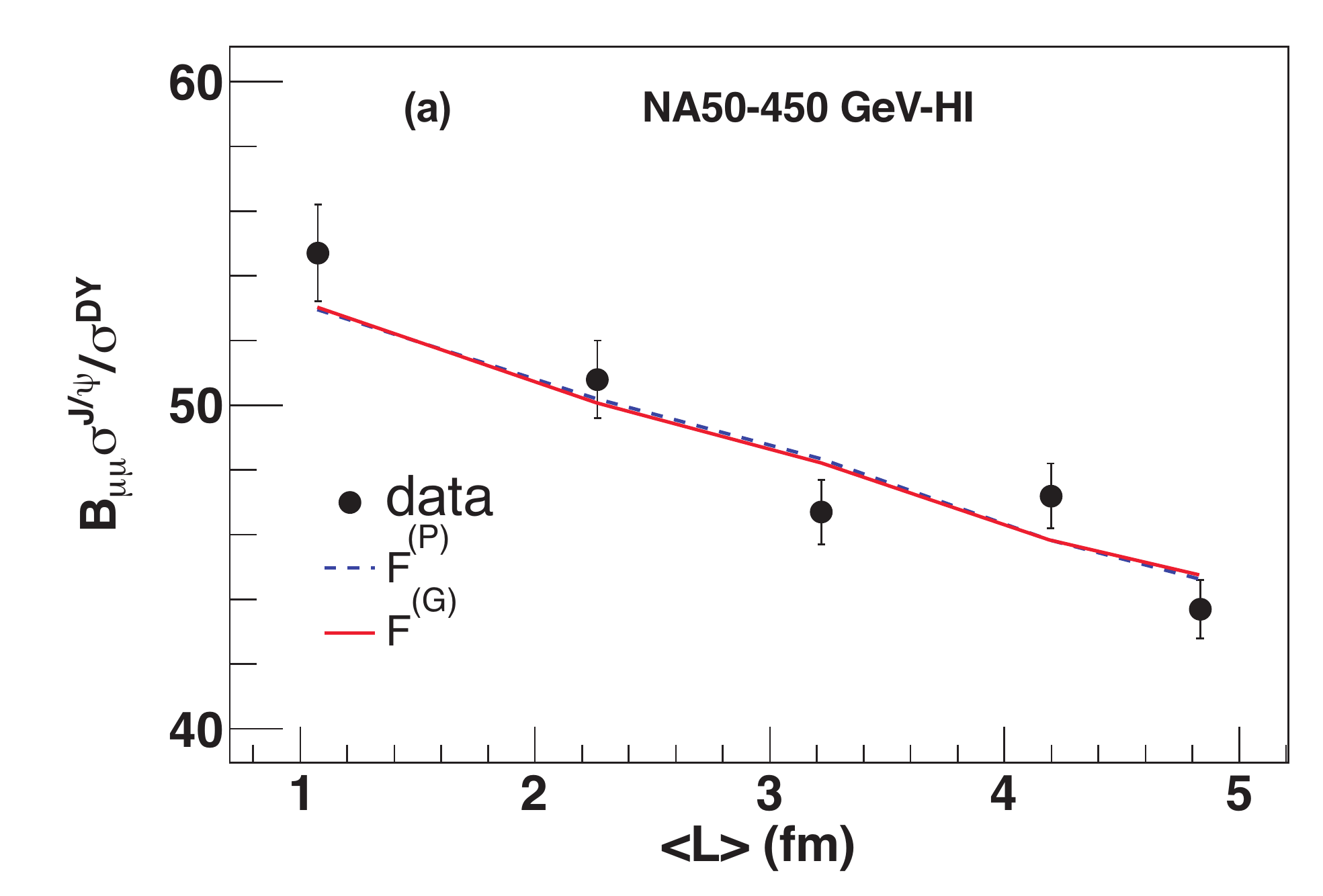}
\includegraphics[width=1.3 \textwidth]{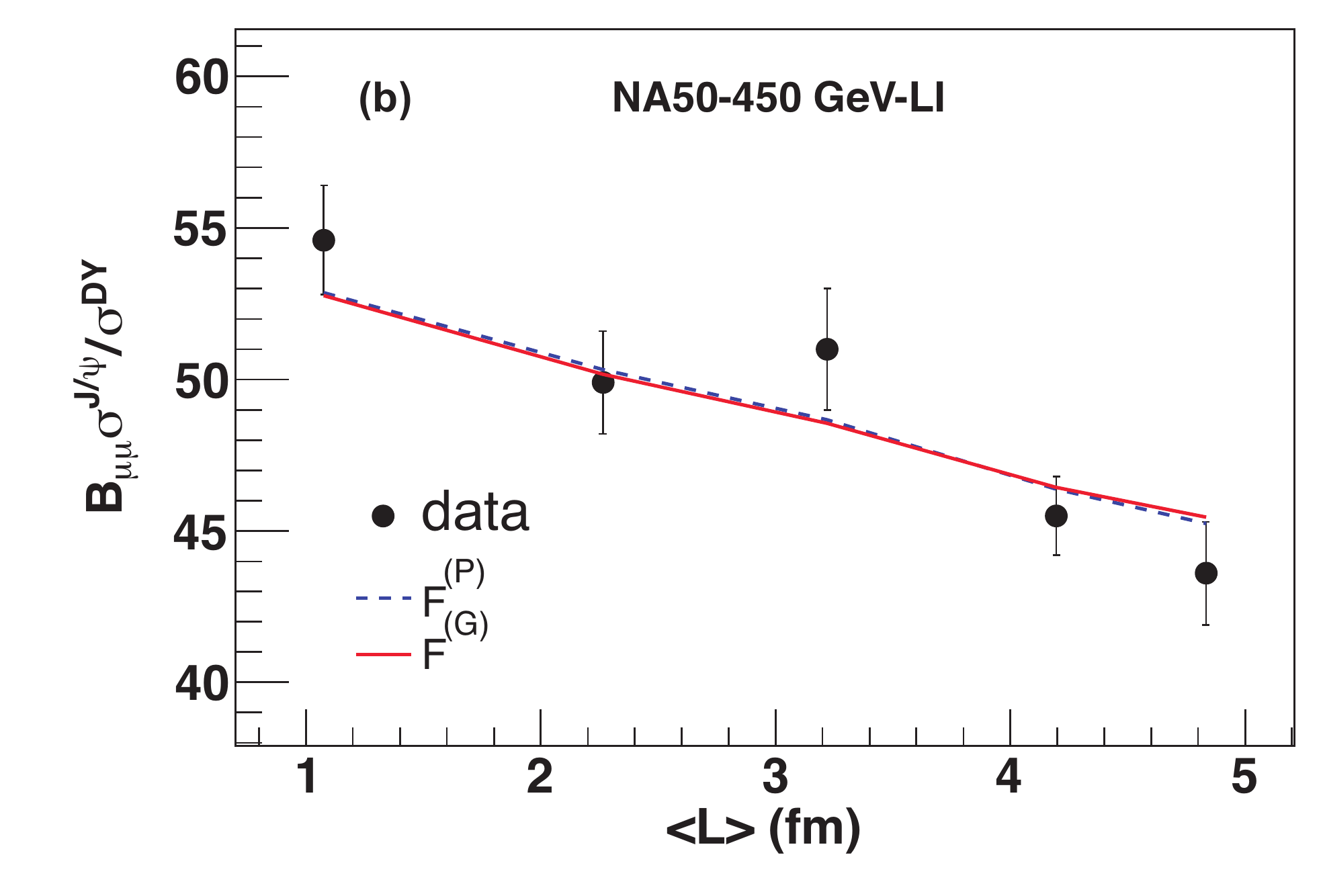}
\includegraphics[width=1.3 \textwidth]{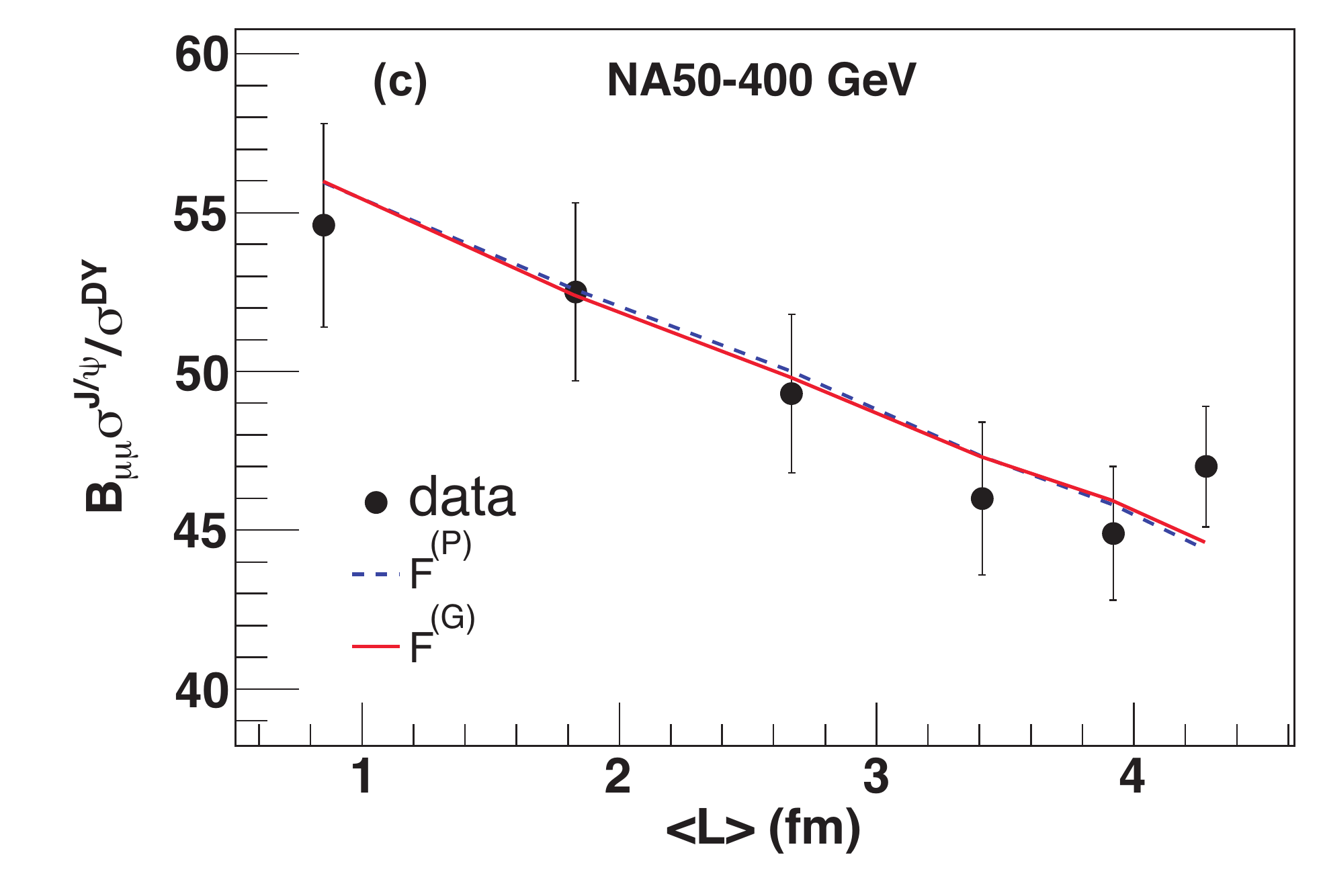}
}
\caption{\footnotesize (Color online) Model description of the data on ratio of the $J/\psi$-to-Drell-Yan production cross-sections in (a) 450 GeV high intensity (HI) run, (b) 450 GeV low intensity (LI) run and (c) 400 GeV $p+A$ collisions, measured by NA50 collaboration at SPS. $B_{\mu\mu}$ denotes the branching ratio of $J/\psi$ decaying into di-muons. The two theoretical curves represents two different parametric forms of $J/\psi$ formation probability ($F(q^{2})$).}
\label{fig1}
\end{figure*}

Organization of the paper is the following. In section II we briefly describe our theoretical framework for calculation of $J/\psi$ and Drell-Yan production cross section in $p+A$ and $A+A$ collisions. Section III is devoted to the analysis of the relevant SPS data. In section IV we give predictions for the centrality dependence of this ratio for the upcoming Compressed Baryonic Matter (CBM) experiment~\cite{CBM} at FAIR. In section V we summarize and conclude. 

\section{Theoretical framework}
%%%%%%%%%%%%%%%%%%%%%%%%%%%%%%%%%%%%%%%%%%%%%%%%%%%%%%%%%%%%%%%%%%%%%%%%%%%%%%%%%%%%
\begin{figure*} 
\scalebox{0.25}
{
\includegraphics[width=1.3 \textwidth]{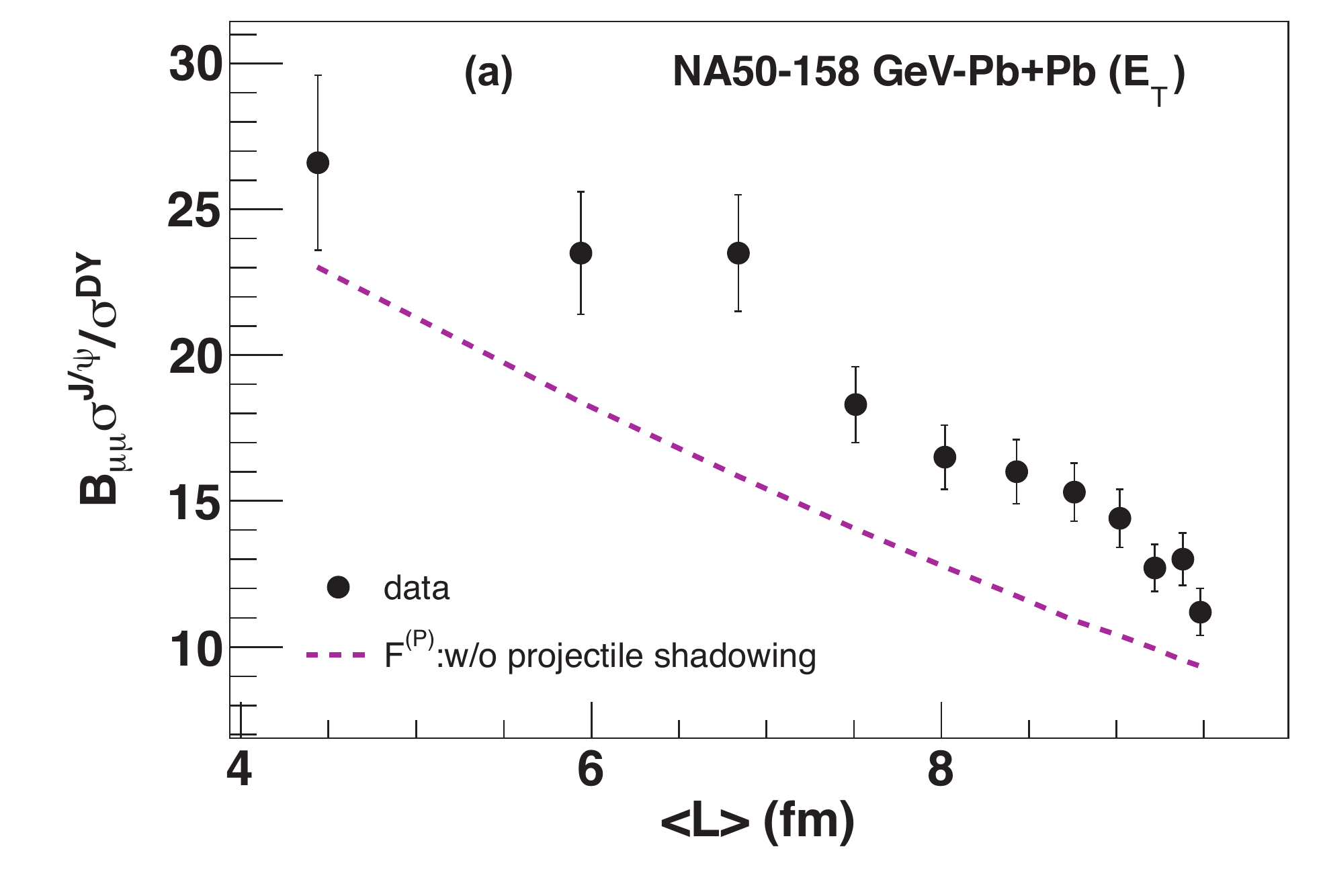}
\includegraphics[width=1.3 \textwidth]{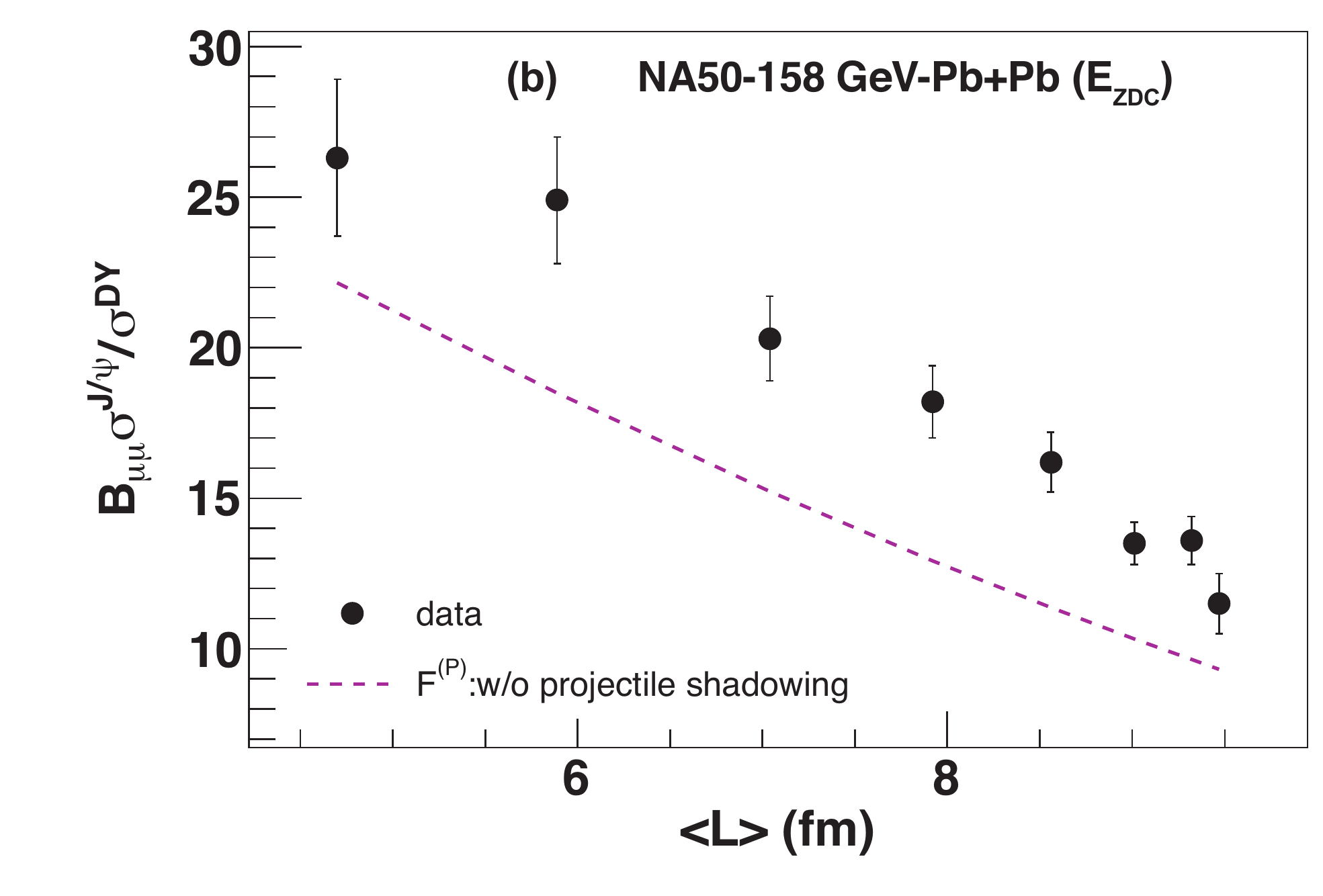}
\includegraphics[width=1.3 \textwidth]{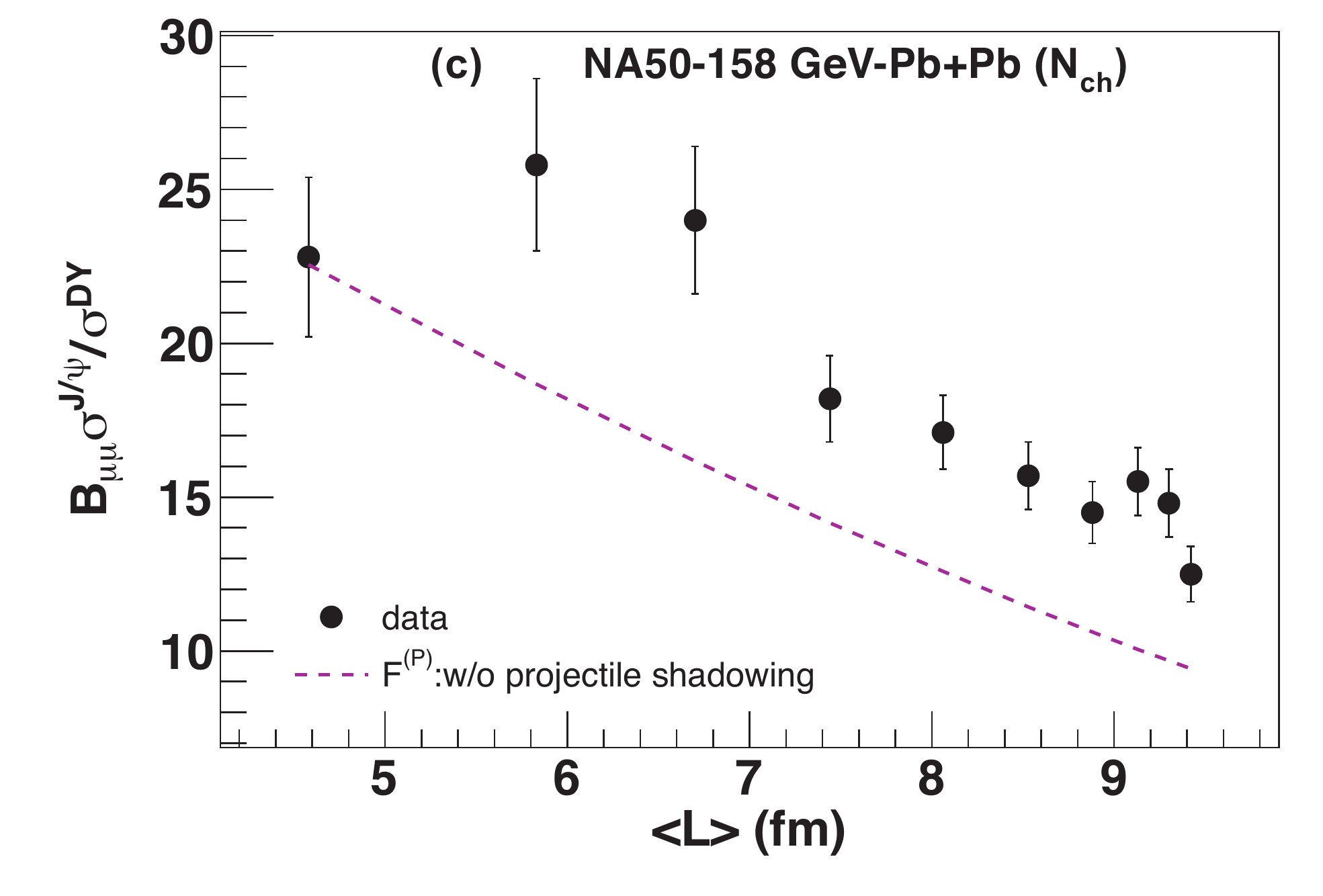}
}
\caption{\footnotesize (Color online) Centrality dependence of $J/\psi$/DY ratio in 158 A GeV Pb+Pb collisions, measured through di-muon channel, within the acceptance ($0 \le y_{c.m.} \le 1$ and $-0.5 \le cos(\theta_{cs}) \le 0.5$) of the NA50 muon spectrometer. Data were published using three independent centrality estimators based on the measurements of : (a) transverse energy ($E_T$), (b) forward energy ($E_{ZDC}$) and (c) total charged particle multiplicity ($N_{ch}$) (see text for details). The theoretical curve correspond to the power law ($F^{(p)} (q^2)$) parametric form of the $J/\psi$ transition probability. Model calculations do not take into account the shadowing corrections of the projectile nucleons and thus completely underestimate the data.}
\label{fig2a}
\end{figure*}

%%%%%%%%%%%%%%%%%%%%%%%%%%%%%%%%%%%%%%%%%%%%%%%%%%%%%%%%%%%%%%%%%%%%%%%%%%%%%%%%%%%%%%%%%%
In order to calculate $J/\psi$ production, we have used the adapted version of the originally proposed QVZ model~\cite{Qui}. Details of the original model and its modifications as used in the present calculations can be found in~\cite{partha1, partha2, Qui,ch02,ch02b}. Here we present a brief description for completeness. The QVZ model assumes $J/\psi$ production hadronic collisions as a factorisable two component process. Production of a $c\bar{c}$ pair, in the initial hard scattering is the first stage and is accounted by perturbative QCD. Hadronization of the $c\bar{c}$ pair to a physical $J/\psi$ meson forms the second stage. This is non-perturbative in nature and can be suitably parametrized following different hadronization schemes. The single differential $J/\psi$ production cross section in collisions of hadrons  $h_1$ and $h_2$, at the center of mass energy $\sqrt{s}$ can be expressed as,
\begin{equation}
\label{diff}
\frac{d\sigma_{h_1h_2}^{J/\psi}}{dx_F} = K_{J/\psi}\int dQ^{2}\left(\frac{d\sigma_{h_1h_2}^{c\bar{c}}}{dQ^2dx_F}\right)\times F_{c\bar{c}
\rightarrow J/\psi}(q^2), 
\end{equation}
\noindent  where $Q^2 = q^2 +4 m_C^2$, with $m_C$ being the charm quark mass and $x_F$ is the Feynman scaling variable. $K_{J/\psi}$ accounts for effective higher order contributions. $F_{c\bar{c} \rightarrow  J/\psi}(q^2)$ denotes the transition probability of a color averaged $c\bar{c}$ pair with relative momentum square $q^2$ to evolve into a physical $J/\psi$ meson, in hadronic collisions. Different parametric forms have been formulated for the transition probability, motivated by the available models of color neutralization. Out of them, two functional forms namely the Gaussian form ($F^{\rm (G)}(q^2)$) and power law form ($F^{\rm (P)}(q^2)$) respectively bearing the essential features of the Color-Singlet~\cite{Singlet} and Color-Octet~\cite{Octet} models have been found earlier to describe the $J/\psi$ production cross section data in $p+A$ collisions reasonably well~\cite{partha1}.
%%%%%%%%%%%%%%%%%%%%%%%%%%%%%%%%%%%%%%%%%%%%%%%%%%%%%%%%%%%%%%%%%%%%%%%%%%%%%%%%%%%%%%%%%%%%%%%%%%%%%%%%%%%%%%%%%%%%%%%%%%%%%%%%%%%%%%%

\begin{figure*} 
\scalebox{0.25}
{
\includegraphics[width=1.3 \textwidth]{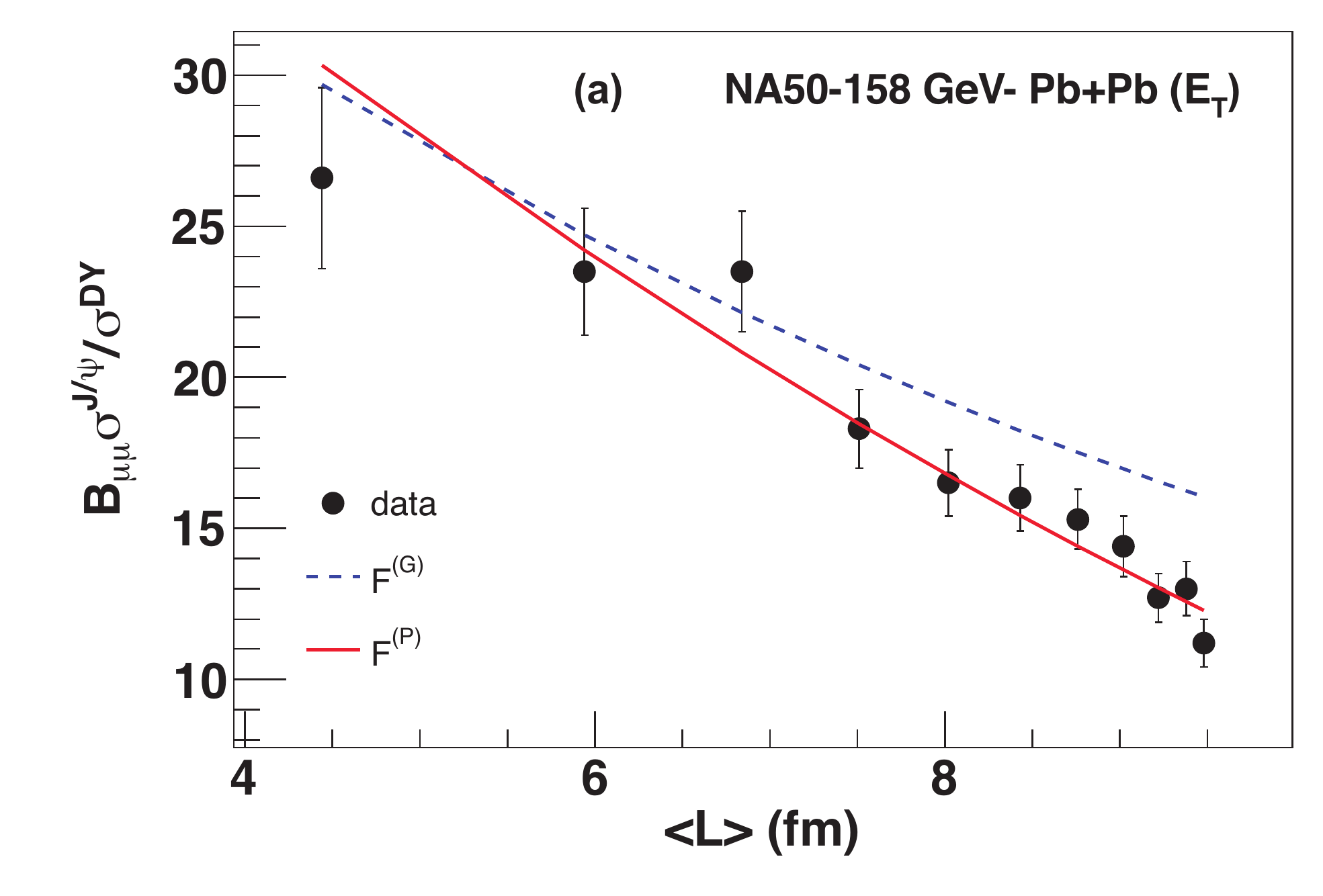}
\includegraphics[width=1.3 \textwidth]{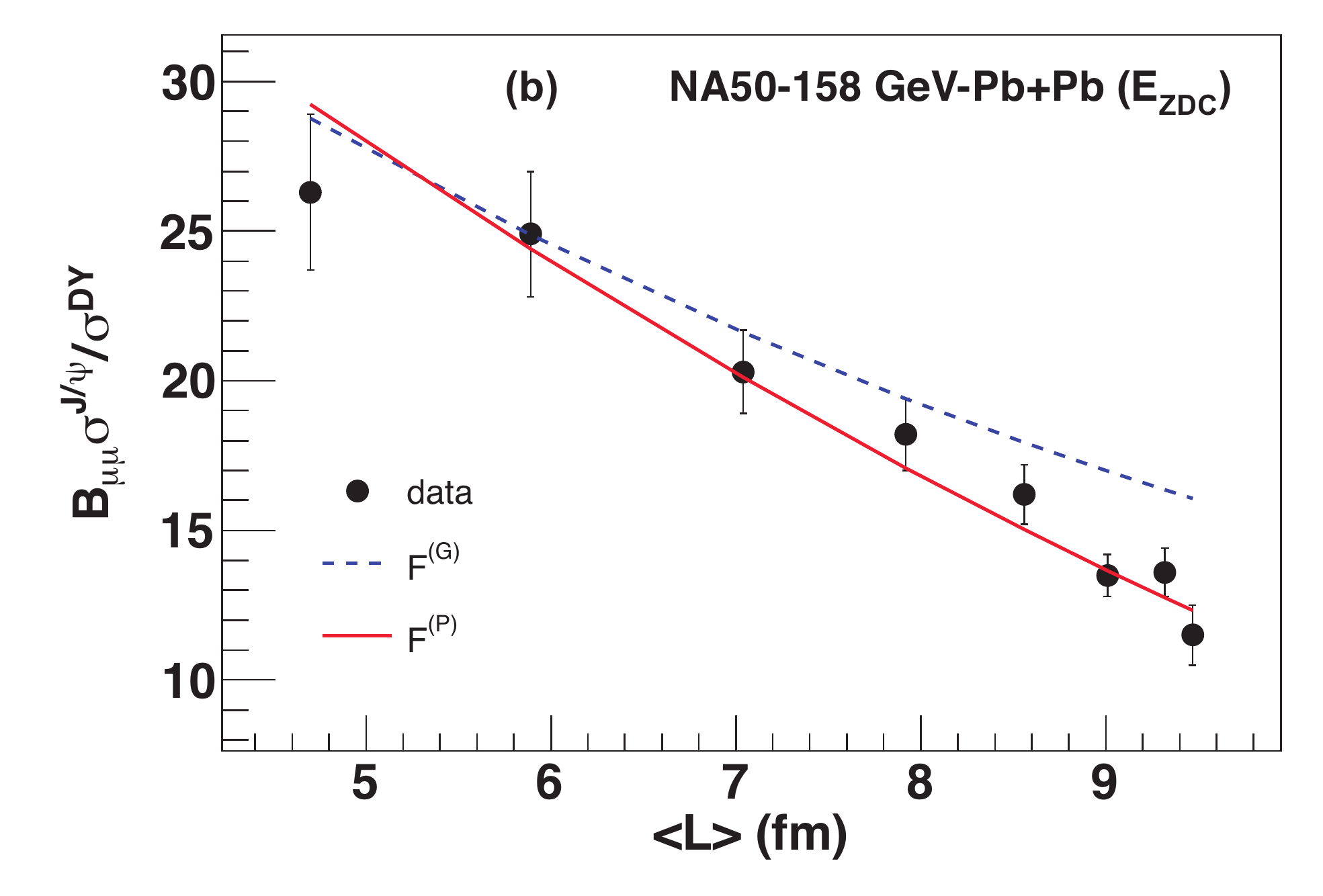}
\includegraphics[width=1.3 \textwidth]{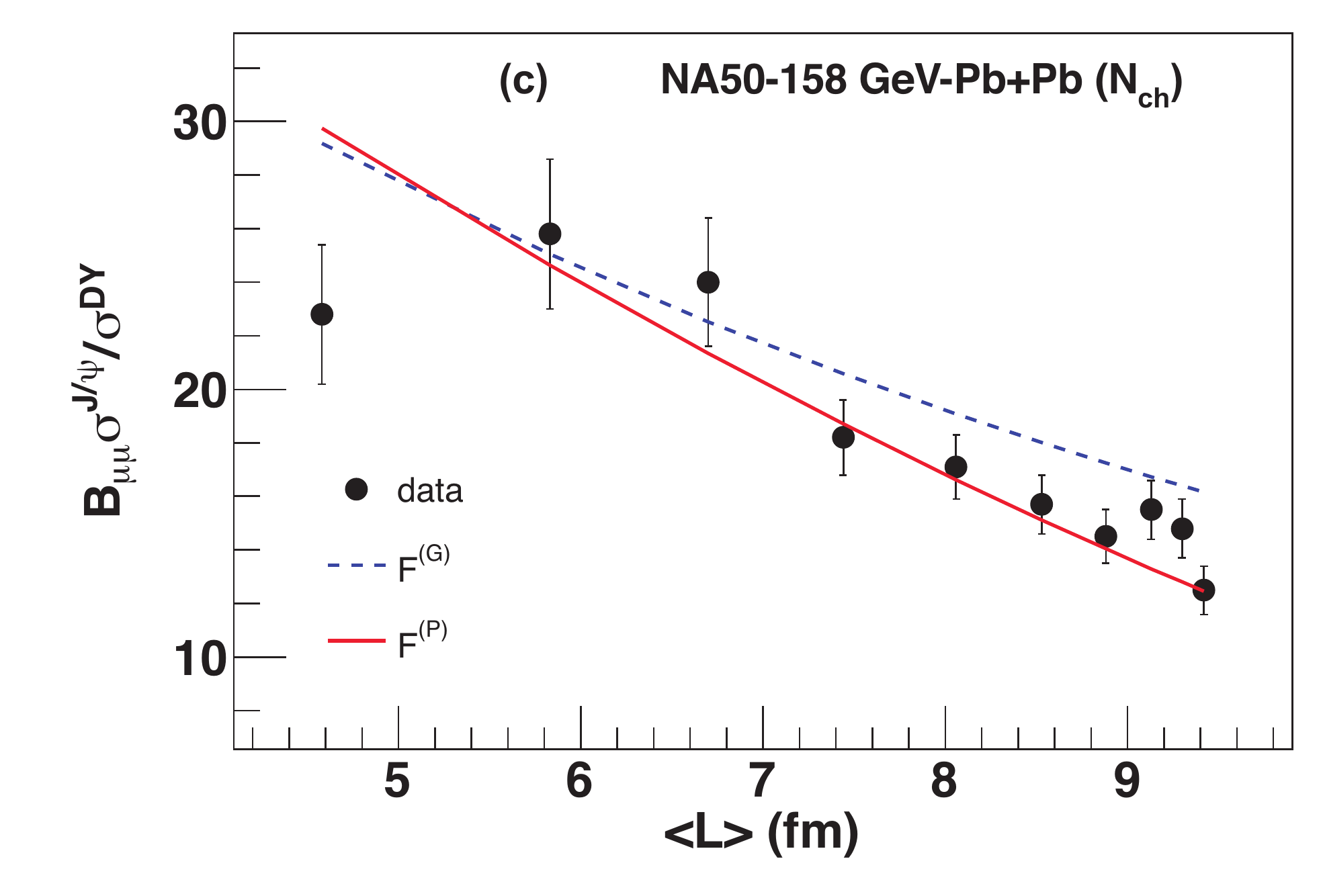}
}
\caption{\footnotesize (Color online) Centrality dependence of $J/\psi$ production in  Pb+Pb  collisions measured at 158 A GeV beam energy. Data analyzed with three independent centrality estimators, namely (a) $E_T$, (b) $E_{ZDC}$ and (c) $N_{ch}$ are shown independently. Data sample were collected within the phase space window: $0 \le y_{c.m.} \le 1$ and $-0.5 \le cos(\theta_{cs})\le 0.5$. The theoretical curves correspond to the two different parametric forms of the $J/\psi$ transition probability. Global shadowing corrections of the projectile and target nucleons are taken into account.}
\label{fig2b}
\end{figure*}

In $p+A$ collisions, $J/\psi$ production is effected by several cold nuclear matter effects that become operative at different stages of evolution. At the initial stage, nuclear modifications of the parton densities of the target nucleons affect the $c\bar{c}$ pair production cross section. The mechanisms governing these modifications are still largely unknown. However based on global DGLAP analysis, several groups have produced parameterizations of the ratio $R_i(A,x,Q^2)$, that convert the free-proton distributions for each parton $i$, $f_i^{p}(x,Q^2)$, into nuclear ones (nPDF), $f_i^A(x,Q^2)$, using the available data from deep inelastic scattering (DIS) and Drell-Yan measurements performed with nuclear targets(see ~\cite{Eskola} for a recent comprehensive review on nPDFS and the references therein). In our analysis, to maintain accordance with our earlier works, we opted for leading order (LO) MSTW2008~\cite{MSTW} set for free proton pdf and LO global EPS09~\cite{EPS09} interface for the ratio $R_i(A,x,Q^2)$.

\begin{figure*} 
\scalebox{0.25}
{
\includegraphics[width=1.3 \textwidth]{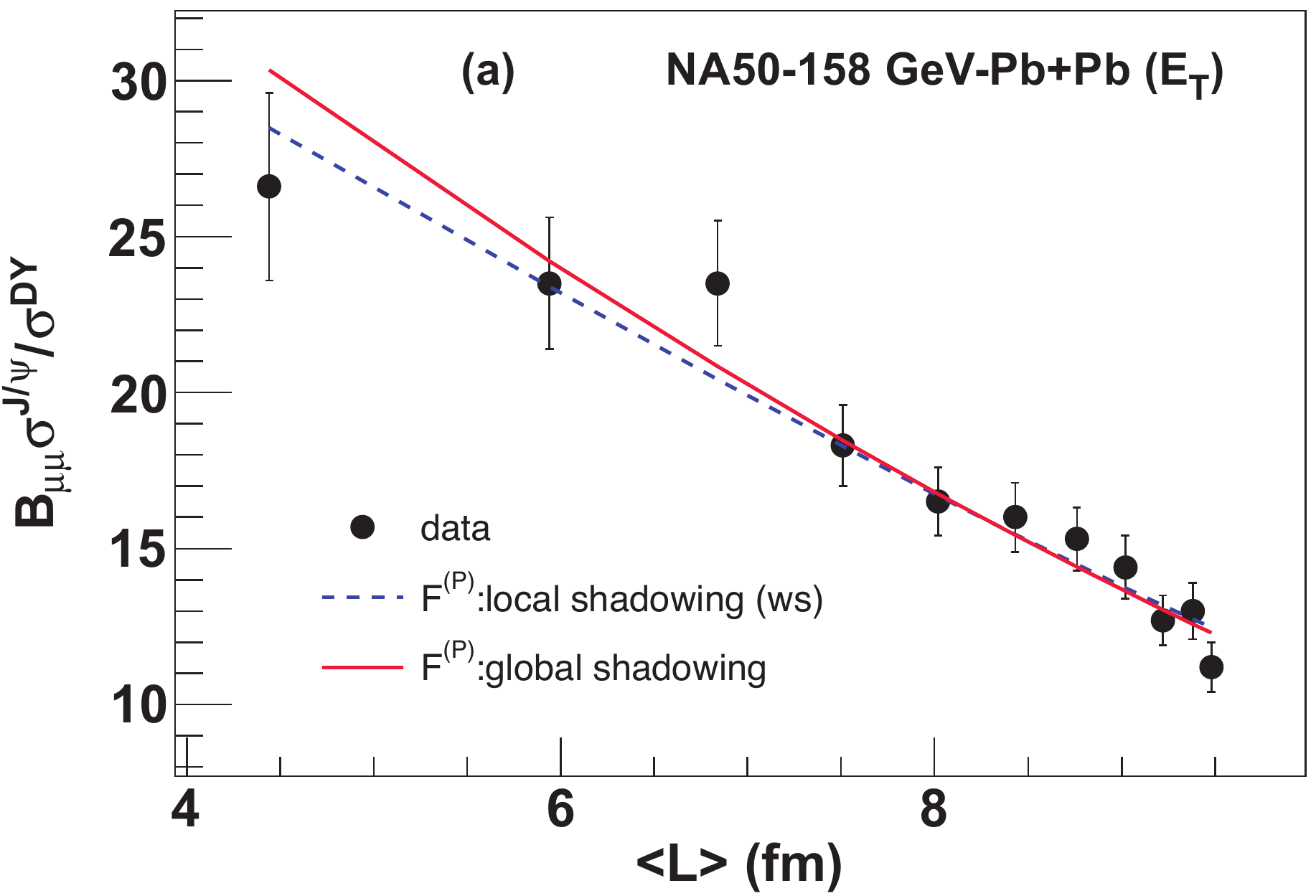}
\includegraphics[width=1.3 \textwidth]{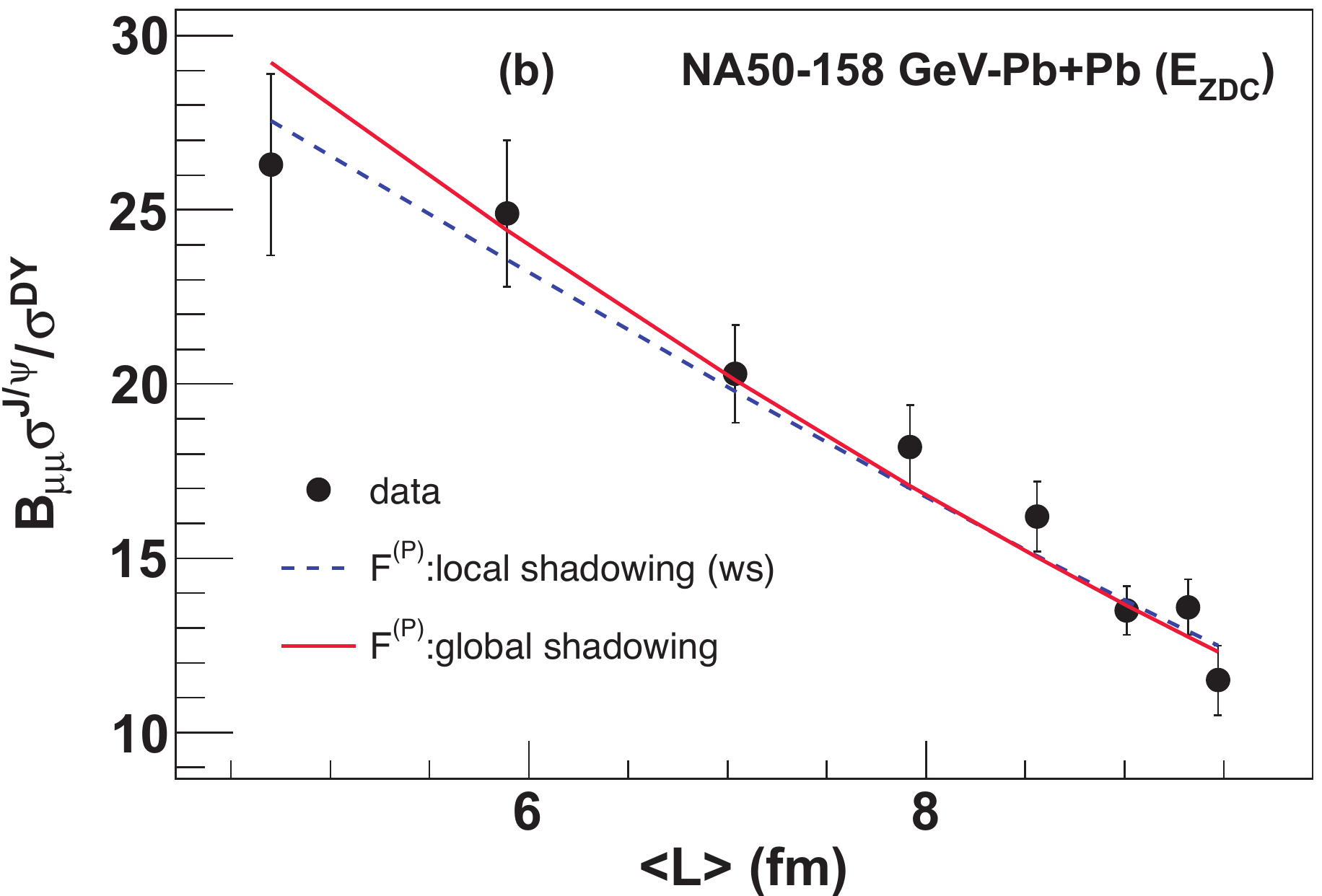}
\includegraphics[width=1.3 \textwidth]{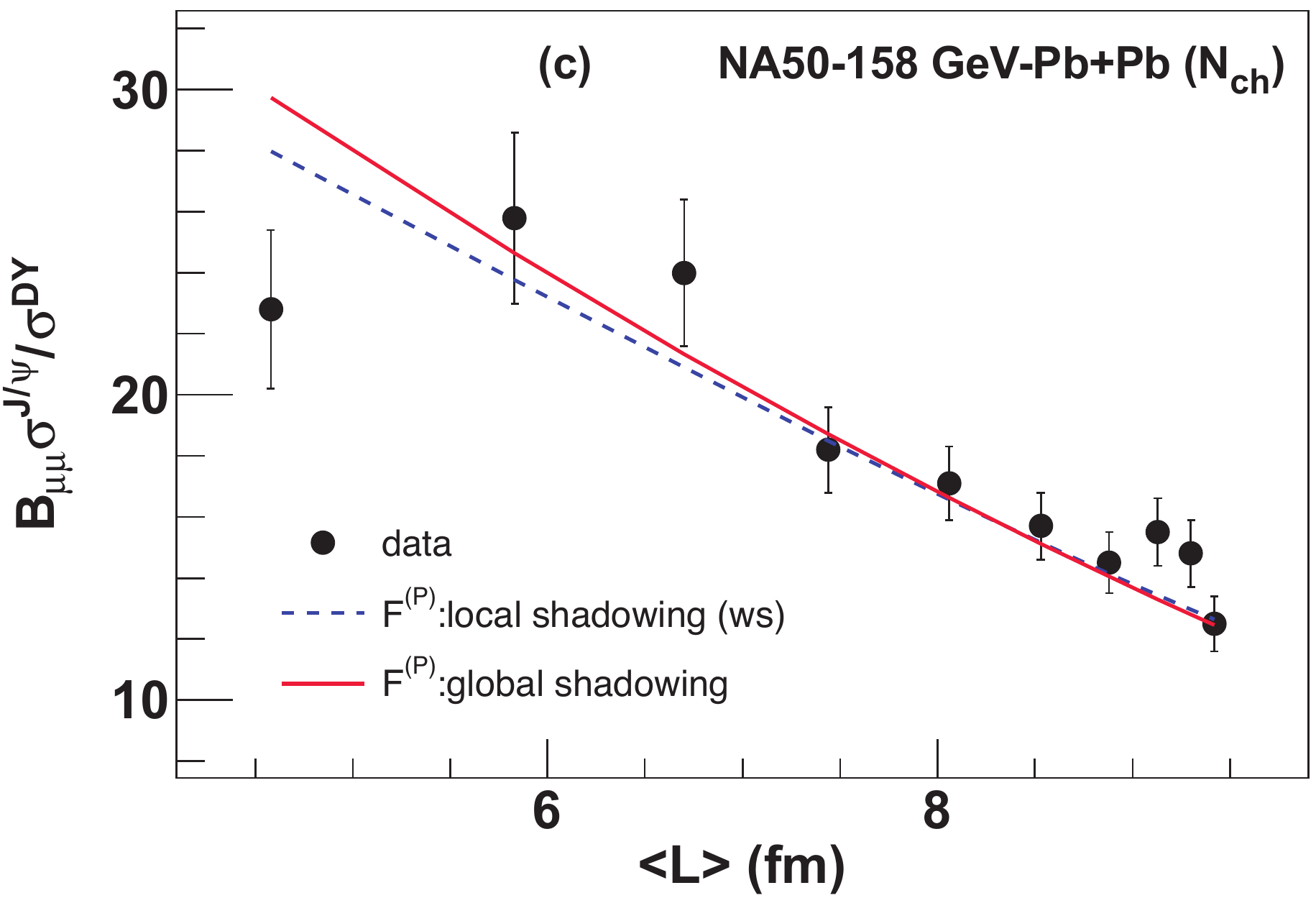}
}

\caption{\footnotesize (Color online) Model calculation of the centrality dependence of $J/\psi$ production  contrasted with the data for Pb+Pb collisions measured by NA50 collaboration at 158 A GeV beam energy and in the kinematic domain: $0 \le y_{c.m.} \le 1$ and $-0.5 \le cos(\theta_{cs}) < 0.5$. Data sets corresponding to all three independent centrality estimators, (a) $E_T$, (b) $E_{ZDC}$ and (c) $N_{ch}$ are shown. Shadowing corrections of the projectile and target nucleons as incorporated in the model, are assumed to be proportional to the local nuclear density. For comparison global shadowing corrections are also shown.}
\label{fig2c}
\end{figure*}

In heavy-ion collisions, parton densities are modified both inside projectile and target nuclei. Depending on the collision geometry, either the halo or the core of the nuclei will be mainly involved in the reaction. Consequently the shadowing effects will be more important in the core than in the periphery. Hence the shadowing factors have to be calculated for various centrality intervals. Different prescriptions are available in literature to model the impact parameter dependent shadowing factors~\cite{Eme99, Kle03, EPS09s}. In the present calculations we implement the spatial dependence of shadowing functions assuming them to be proportional to the local nuclear density~\cite{Eme99, Kle03}, with Woods-Saxon (WS) profile chosen for nuclear density distributions. Inhomogeneous shadowing effects inside the nucleus can also be invoked by postulating them to be linearly proportional to the density weighted longitudinal thickness of nucleus at the transeverse position of the binary collision ($T_A ({\bf r_T}$) defined as  $T_A ({\bf r_T})=\int \rho_A ({\bf{r_T}},z) dz$, where $\rho_A ({\bf{r_T}},z)$ is the local nuclear density at a point $({\bf{r_T}},z)$ inside the nucleus $A$. However these two parameterizations of local shadowing have been found to give similar result at SPS energies, their difference lies within $2 \%-3 \%$~\cite{shadow2}. Recently spatial dependence of the nPDFs has been studied in detail using the $A$-dependence of the spatially independent global EPS09 and EKS98 routines~\cite{EPS09s}. Spatial dependence is modelled as a power series of $T_{A}$, having terms up to $(T_A)^4$. Two spatially dependent nPDF sets namely EPS09s and EKS98s have also been released for public use. 

In $p+A$ and $A+A$ collisions, the nascent $c\bar{c}$ pairs once produced will pass through the nuclear medium. During their passage they undergo multiple collisions with the medium and gain relative square four momentum at a rate of $\varepsilon^2$ per unit path length. Some of the $c\bar{c}$ pairs can thus gain enough momentum to cross the threshold to evolve as open charm mesons ($D\bar{D}$ pairs). This would finally result in the reduction of $J/\psi$ yield compared to the nucleon-nucleon collisions. The overall effect of the multiple scattering of the $c\bar{c}$ pairs can be represented by a shift of $q^2$ in the transition probability, 
\begin{equation}
q^2  \longrightarrow  \bar{q}^2 = q^2 + \varepsilon^2\, <L> \ .
\label{q2shift}
\end{equation}
 where, $<L>$ denotes the average geometrical path length traversed by the $c\bar{c}$ pair till it exits the medium and is calculated using Glauber model. For both parameterizations of transition probability, $F(q^2)$, the corresponding values of $\varepsilon^2$, extracted from the analysis of absolute $J/\psi$ production cross-sections in $p+A$ collisions, exhibited non-trivial dependence on the beam energy ($E_b$)~\cite{partha1}. Lower be the beam energy, larger is the value of $\varepsilon^2$, implying larger nuclear dissociation.

In Drell-Yan process, a quark and an anti-quark from the nucleons of the two colliding nuclei annihilate to form a virtual photon which subsequently decays into a $\mu^{+}\mu^{-}$ pair. The leading order Drell-Yan cross section can be calculated by using the standard prescription (see for example, Ref.~\cite{Vogt}). They are not affected by any final state effect. Only CNM effect incorporated is the nuclear modification of the parton distribution functions. The inclusive cross sections can be obtained by integrating the double differential cross section within the suitable mass and rapidity range as appropriate for a particular experimental set up.  

\section{Analysis of SPS data}

\begin{figure*} 
\scalebox{0.25}
{
\includegraphics[width=1.3 \textwidth]{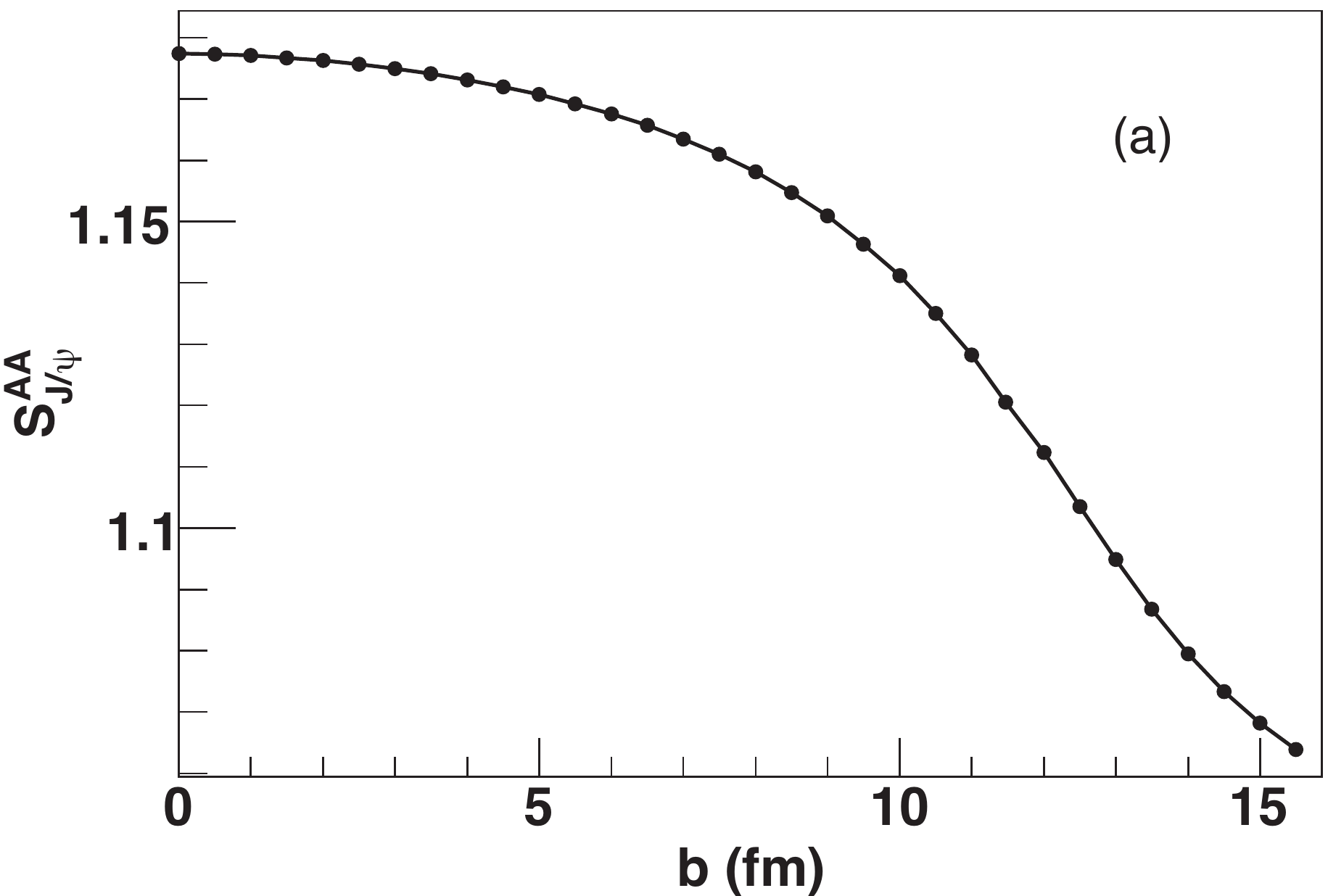}
\includegraphics[width=1.3 \textwidth]{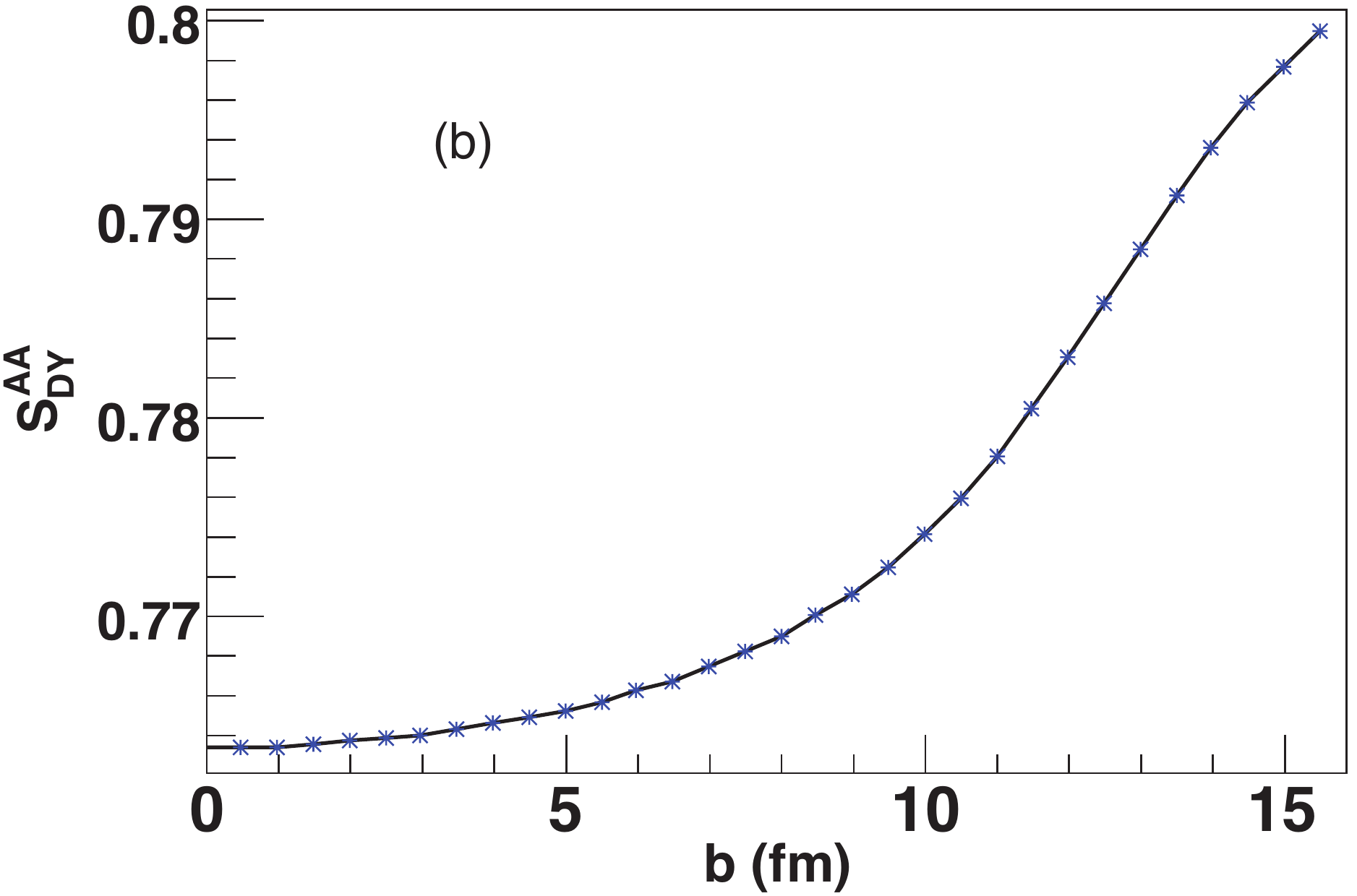}
\includegraphics[width=1.3 \textwidth]{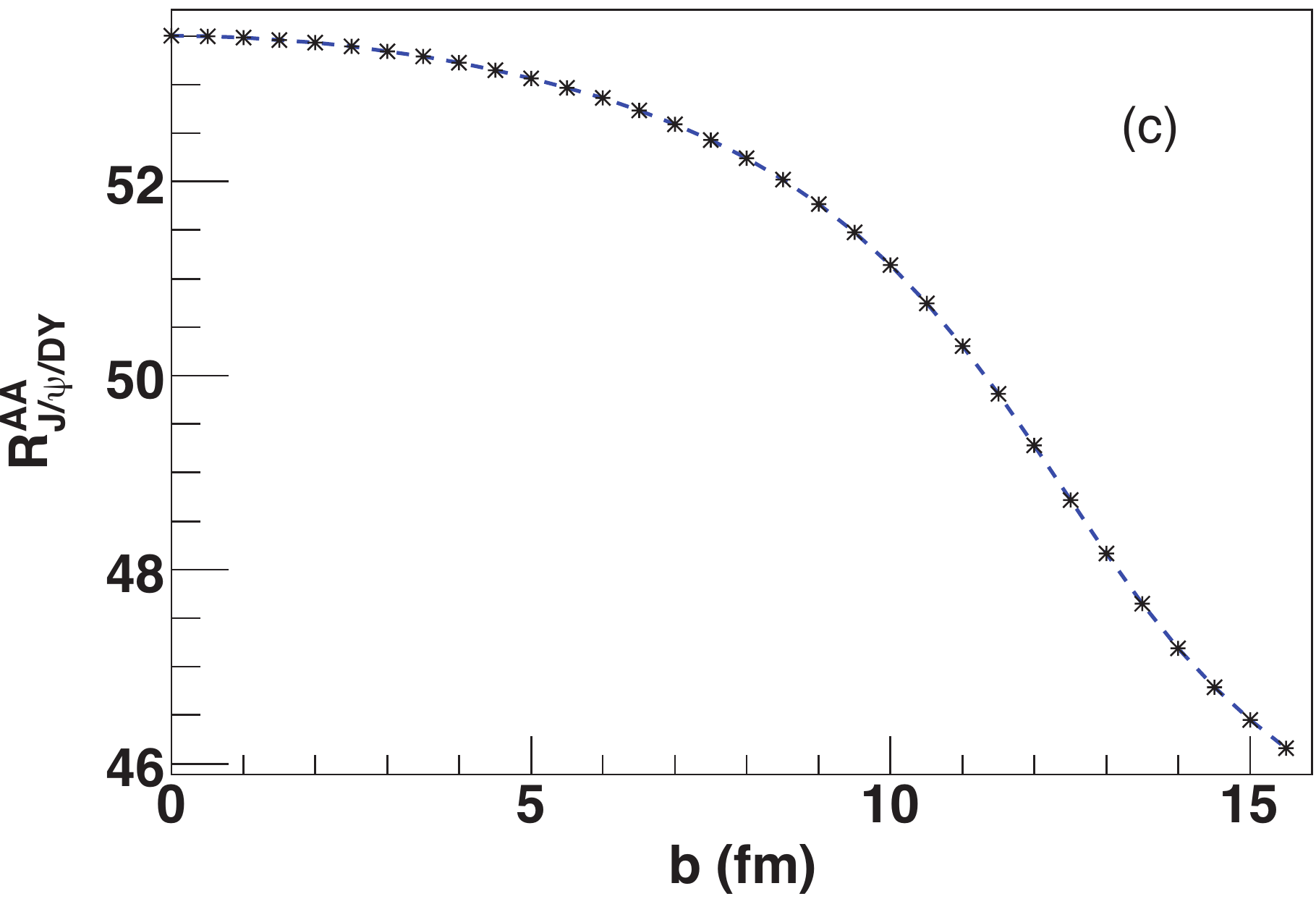}
}
\caption{\footnotesize (Color online) Centrality dependence of the shadowing function for (a) $J/\psi$ (left), (b) Drell-Yan productions (middle) and (c) ratio of the two production cross sections in di-muon channel, in absence of any final state nuclear dissociation of $J/\psi$ ($R^{AA}_{J/\psi/DY}=B_{\mu\mu}\sigma^{AA}_{J/\psi}(\varepsilon^2=0)/\sigma^{AA}_{DY}$) (right), estimated for 158 A GeV Pb+Pb collisions. Total cross-sections are obtained by integrating the differential cross-section in the rapidity range $0 < y_{c.m.} < 1$, which corresponds to the rapidity coverage of the NA50 di-muon system for the Pb-Pb run.}
\label{fig3}
\end{figure*}

With this brief description of the model, we will now move forward to examine the reliability of the model calculations in describing the available data sets on $J/\psi$-to-Drell-Yan ratio at SPS, available from NA50 collaboration. Uncertainties in the ratio are mostly dominated by the statistical errors of the Drell-Yan production cross-section data. NA50 measured $J/\psi$ through its di-muon decay channel. For $p+A$ collisions the said ratio was measured with both 400 GeV~\cite{NA50-400} and 450 GeV~\cite{NA50-450} proton beams. In case of Pb+Pb collisions the data were collected at 158 A GeV~\cite{Ale05}. For 450 GeV, NA50 collected two sets of $p+A$ data in two independent runs with different beam intensities. The first (ever) set were collected with a high intensity (HI) 450 GeV proton beam using five different nuclear targets ($Be, Al, Cu, Ag, W$). Subsequently at the same beam energy new $p+A$ data samples were collected with low beam intensity (LI) and with same set of targets. The data sets at 450 GeV are corresponded to the phase space window: $-0.5 < y_{c.m.} <0.5$ and $-0.5 < cos(\theta_{cs}) < 0.5$, where $y_{c.m.}$ denotes di-muon centre-of-mass rapidity and $\theta_{cs}$ is the Collins-Soper angle. NA50 also took data with 400 GeV incident proton beam using six different target nuclei ($Be, Al, Cu, Ag, W, Pb$). Data were selected in the kinemtaic region: $-0.425 < y_{c.m.} < 0.575 $ and $-0.5 < cos(\theta_{cs}) < 0.5$. Unlike previous measurements, in 400 GeV run, data for all targets were collected in the same data taking period which siginificantly reduces the systematic errors. 

In Fig.~\ref{fig1} we show the variation of the $J/\psi$-to-Drell-Yan cross section ratio as a function $<L>$, for various $p+A$ systems. The two theory curves result from fitting the above data sets following two parameterizations of $F(q^2)$, representing two different physical mechanisms of $J/\psi$ formation, as mentioned earlier. For both the cases nuclear modification of parton densities are taken into account. The $<L>$ values as published with the data are used to generate the theoretical curves. Different $<L>$ correspond to different target nuclei, for which corresponding EPS09 set of nPDF is implemented. Note that for a given parameterization of $F(q^2)$ (power-law or Gaussian), our employed model contains three parameters. Two of them, namely $\epsilon^2$ and $\alpha_F$  were already fixed earlier~\cite{partha1} using inclusive data on absolute $J/\psi$ production cross sections in $p+A$ and $p+p$ collisions respectively. The only free parameter left for the current analysis is the $K_{eff}$ defined as $K_{eff}=f_{J/\psi}/K_{DY}$, where $f_{J/\psi} = K_{J/\psi} \times N_{J/\psi}$. $K_{DY}$ takes care of higher order effects in Drell-Yan production. As evident from the figure, both the curves can give a satisfactory description of the data. This is of course expected from our earlier observation, where we found that the $J/\psi$ production in $p+A$ collisions at SPS energies is well described by both forms of $F(q^{2})$.

\begin{figure} 
\includegraphics[height=5.5cm,width=6.5cm]{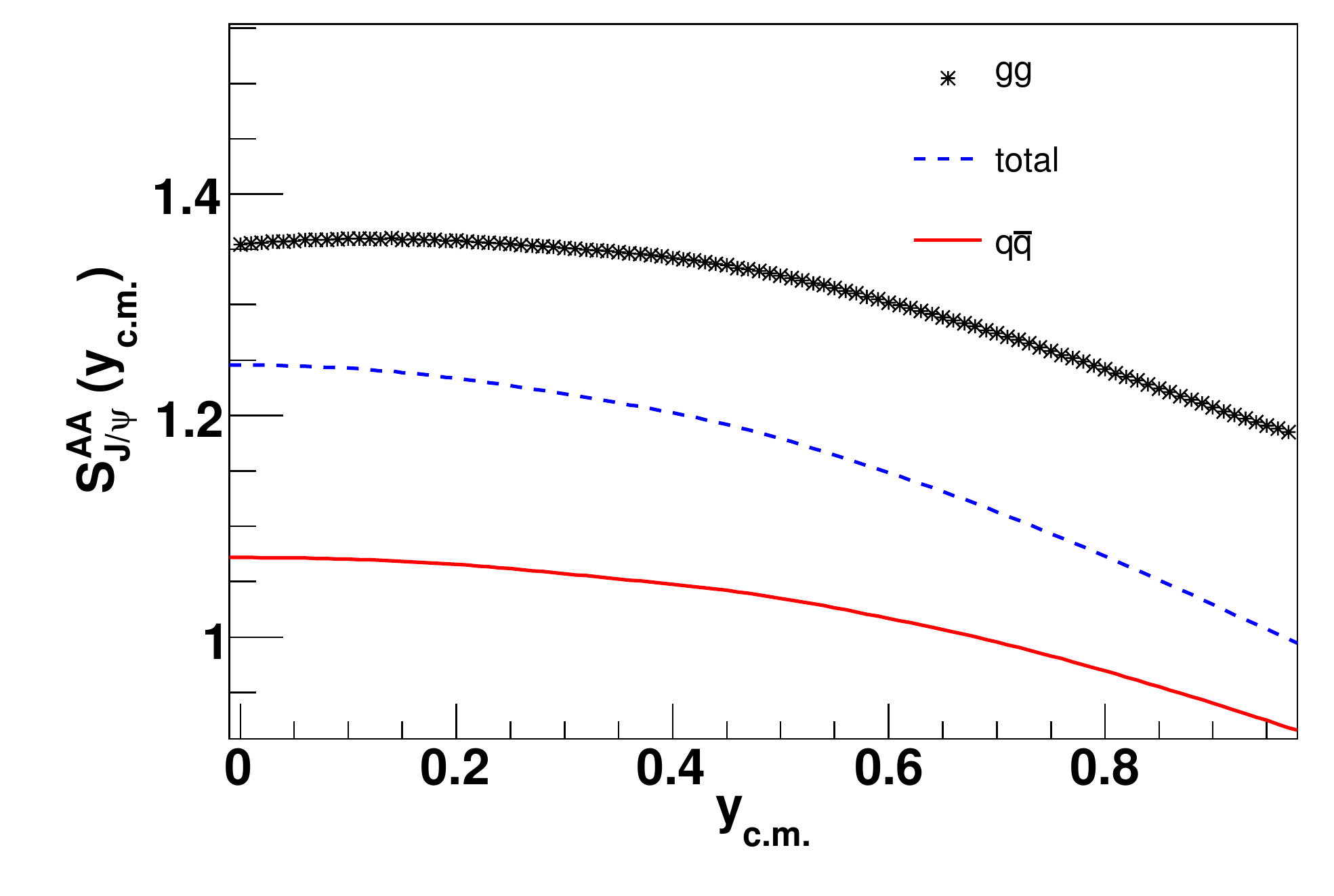}
\caption{\footnotesize (Color online) Rapidity dependence of the shadowing function for $J/\psi$ in 158 A GeV Pb+Pb collisions within the coverage of NA50 muon system. The contributions from quark-antiquark ($q\bar{q}$) annihilation and gluon-gluon ($gg$) fusion are shown separately along with the total cross section.}
\label{fig3a}
\end{figure}

Let us now examine the ratio in Pb+Pb collisions at 158 A GeV. In this case, all the model parameters, as mentioned earlier, are fixed from the $p+A$ data. We have selected the latest data set published by NA50 collaboration~\cite{Ale05}. %The data sample corresponded to their new measurement of $J/\psi$ production in year 2000, under improved experimental conditions with respect to previous measurements. The target system was placed in a vacuum chamber and the setup was better adapted to study in particular the most peripheral nuclear collisions with unprecedented accuracy, which was otherwise polluted by out-of-target interactions particularly Pb-air interactions. 
Data were collected in the di-muon kinematic domain $0<y_{c.m.}<1$ and $-0.5 < cos(\theta_{cs}) <0.5$. Drell-Yan differential cross-sections are integrated in the mass domain $2.9 - 4.5$ GeV/$c^2$. Analysis has been performed using three independent centrality estimators, namely, neutral transverse energy ($E_T$) deposited in electromagnetic calorimeter, forward energy ($E_{ZDC}$) deposited in zero degree calorimeter and charge particle multiplicity per unit of pseudo-rapidity at mid-rapidity $(dN_{ch}/d\eta)_{max}$ (designated by $N_{ch}$ in the figures). For each case, the centrality classes and the corresponding values of number of participant nucleons ($N_{part}$), of impact parameter ($b$), and of the average path length of nuclear matter traversed by the pre-resonant $c\bar{c}$ pair ($<L>$) are also given. As earlier, we have used those $<L>$ values in our model calculations. It is also important to mention that for Pb+Pb collisions, no model parameter is tuned. All the model parameters are fixed from the present ($k_{eff}$) and previous ($\alpha_F$ and $\epsilon^2$) analysis of the available $p+A$ data sets.
\begin{figure*} 
\scalebox{0.25}
{
\includegraphics[width=1.3 \textwidth]{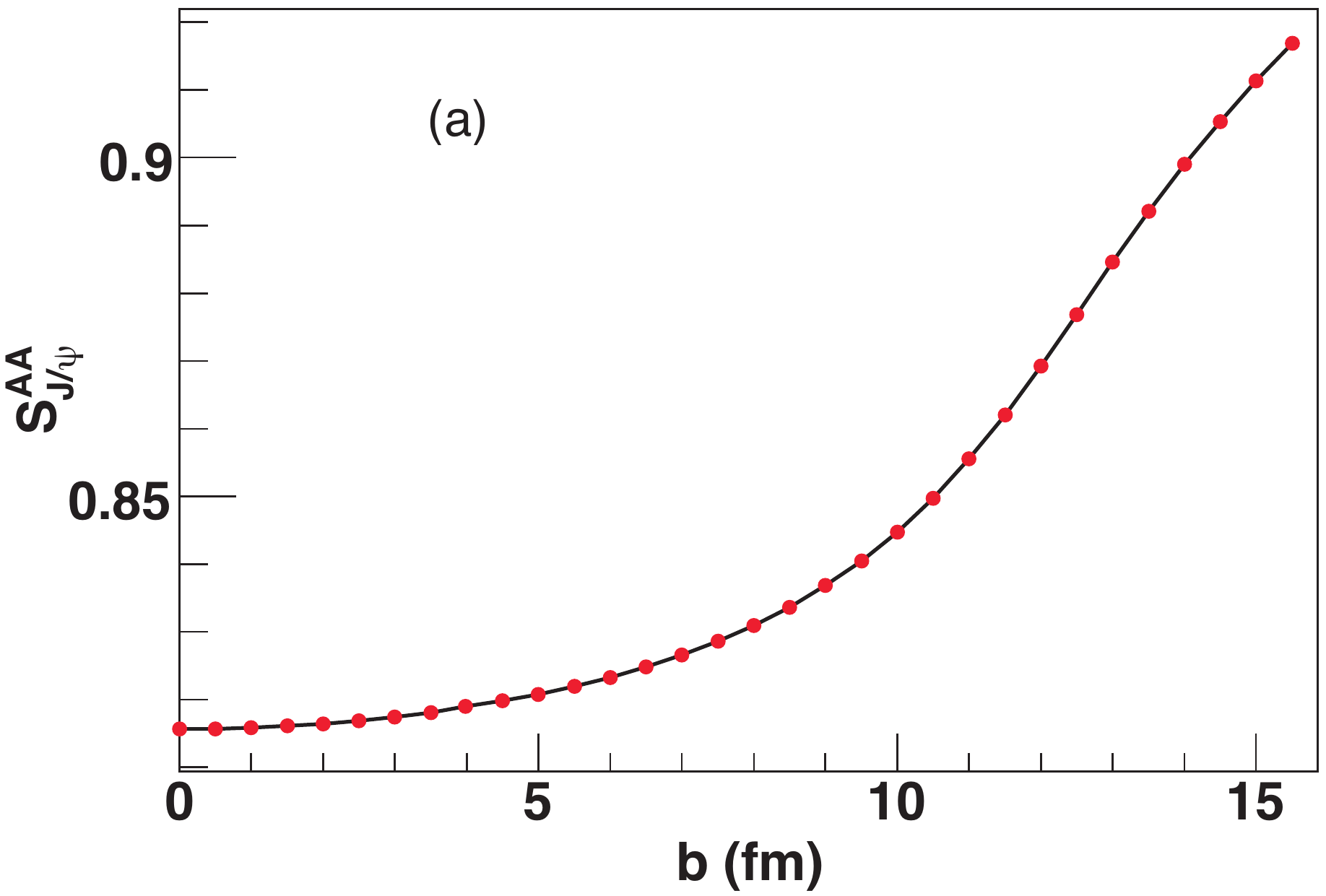}
\hspace{5pt}
\includegraphics[width=1.3 \textwidth]{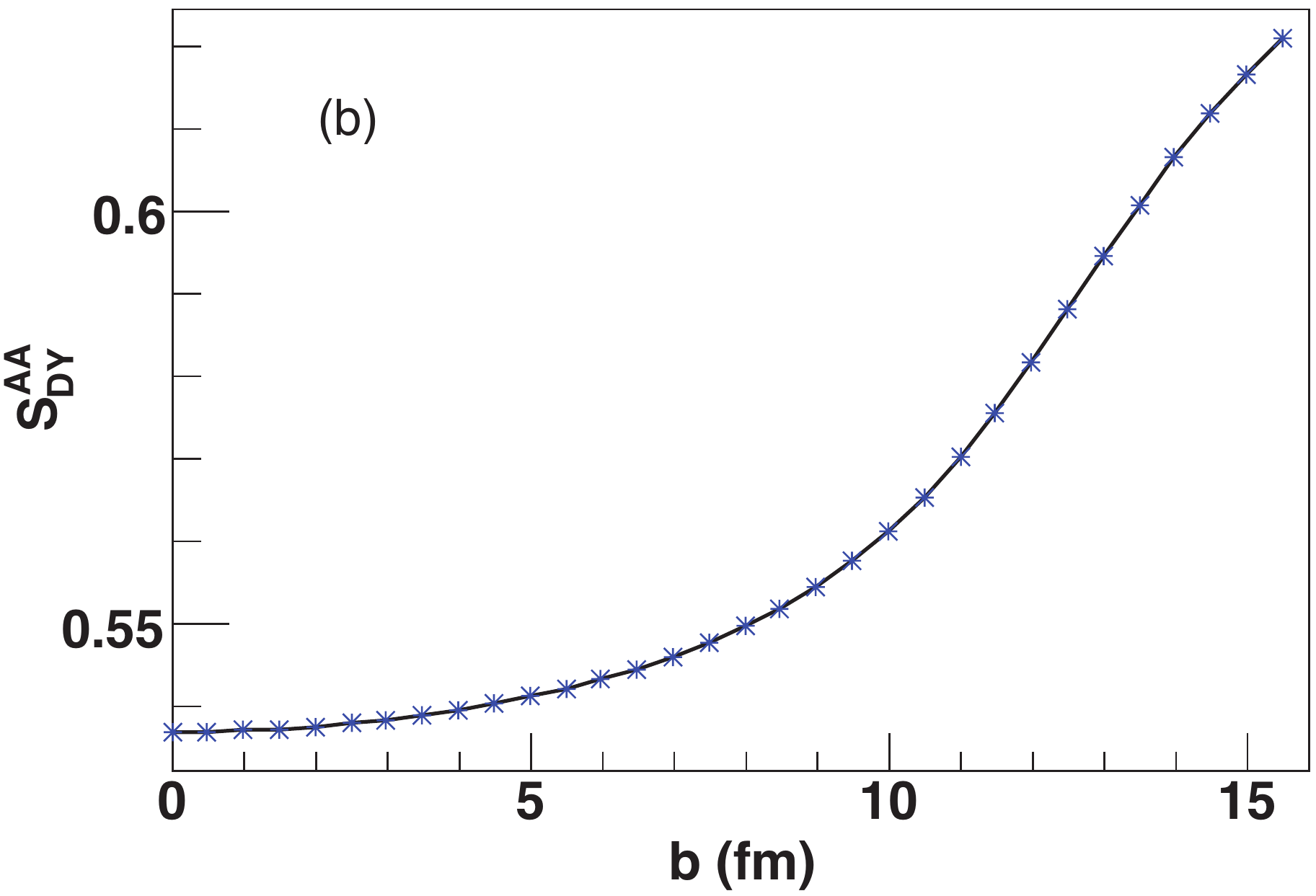}
\hspace{5pt}
\includegraphics[width=1.3 \textwidth]{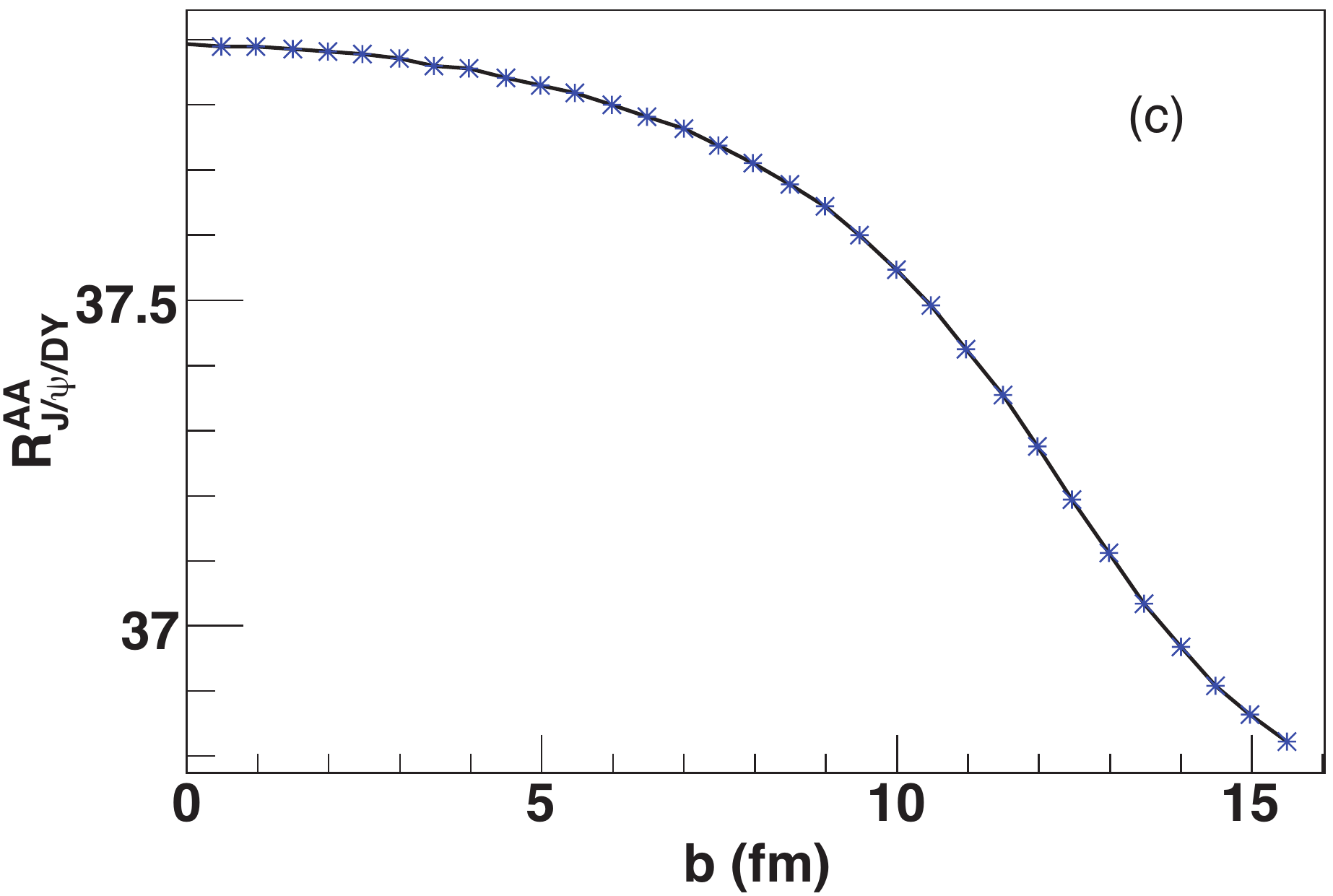}
}
\caption{\footnotesize (Color online) Impact parameter dependence of the shadowing functions, for (a) $J/\psi$ (left), (b) Drell-Yan productions (middle) and (c) the ratio of the two production cross sections in absence of any final state interaction of $J/\psi$ in the nuclear medium (right). Shadowing factors are evaluated at mid-rapidity, in 25 A GeV Au+Au collisions.}
\label{fig4}
\end{figure*}

To investigate the role played by parton shadowing in detail let us first compare the data with our model calculations without incorporating the nuclear modifications of projectile nucleons. Results are illustrated in Fig.~\ref{fig2a}. The theoretical curves correspond to the power law parametrization of $F(q^2)$. As evident from the figure, the theoretical curves completely underestimate the data (Same is true for the Gaussian though not shown explicitly). The dominant contribution to $J/\psi$ production comes from gluon fusion and at the SPS kinematic domain gluons in the projectile nucleons are heavily anti-shadowed, within EPS09 routines. Once these shadowing corrections of projectile partons are taken into account, a much better description of the data is obtained as seen in Fig.~\ref{fig2b}. Here we have contrasted the data with both forms of $F(q^2)$. For both the cases we have considered global shadowing implying the shadowing factors are independent of the spatial location of the partons. In consistency with the previous observations, power-law form ($F^{(P)}(q^2)$), due to threshold effects, generates much larger suppression compared to the Gaussian ($F^{(G)}(q^2)$) case and thus can explain the Pb+Pb data even for most central collisions. Gaussian form on the other hand underestimates the suppression at least for higher centralities. Note that suppression for $F^{(G)}(q^2)$ is equivalent to that generated by Glauber model with Eikonal approximation~\cite{partha1}. Finally in Fig.~\ref{fig2c}, we have examined the effects of density dependent local shadowing as implemented in the model. The effect of spatial dependence of shadowing is more visible for peripheral collisions. As stated earlier, at the SPS energy regime gluon densities in both target and projectile exhibit over populations and hence when a spatial inhomogenity is incorporated, gluons at the core are more effected compared to those near the surface. Hence in peripheral collisions the $J/\psi$ production cross section is less for local shadowing compared to the global case (where the amount of anti-shadowing is independent of spatial position of the partons and thus same for different centrality intervals) and for most central collision it is more. To examine such geometrical effects in more detail, let us now calculate the shadowing function $({S_{J/\psi}}(S_{DY}))$, defined as the ratio between $J/\psi$ (or Drell-Yan) production cross-sections in Pb+Pb and p+p collisions in absence of any final state interaction. Owing to difference in quark and gluon shadowing effects, behavior of the resulting shadowing functions for $J/\psi$ and Drell-Yan productions will also be different. This is illustrated in Fig.~\ref{fig3}, where we have plotted the variation of the shadowing functions in 158 A GeV Pb+Pb collisions as a function of $b$.

\begin{figure} 
%\centering
\scalebox{1}
{
\hspace{-30pt}
\includegraphics[width=0.4\textwidth]{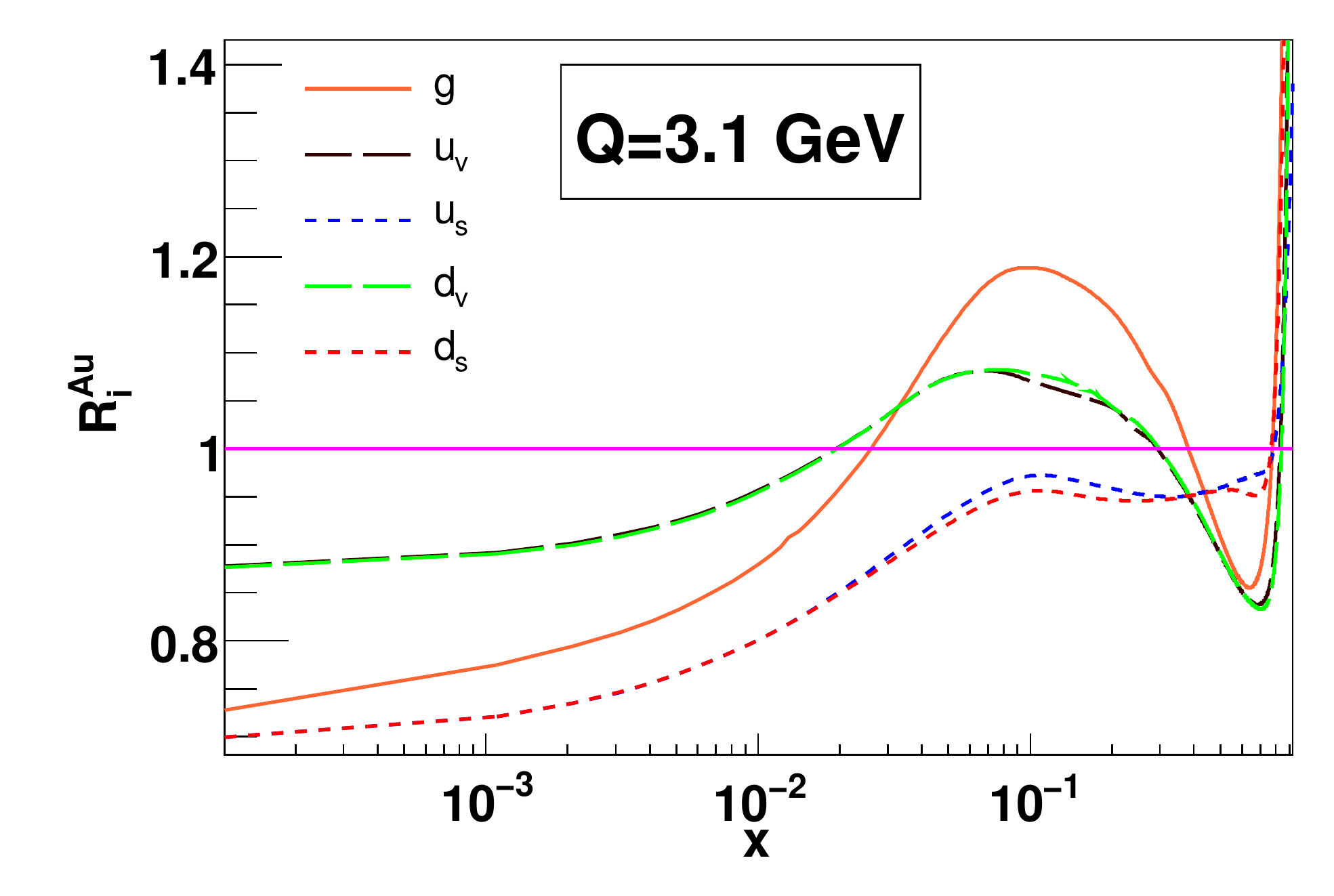}
}
\caption{\footnotesize (Color online) Variation of the nuclear parton distribution (nPDF), inside Au nucleus, as a function of the momentum fraction $x$. Parton densities are evaluated following LO EPS09 default parametrization at a momentum scale $Q=3.1$ GeV, appropriate for $J/\psi$ production. The uncertainties in the nPDF sets are not taken into consideration.}
\label{fig5}
\end{figure}

%%%%%%%%%%%%%%%%%%%%%%%%%%%%%%%%%%%%%%%%%%%%%%%%%%%%%%%%%%%%%%%%%%%%%%%%%%%%%%%%%%%%%%%%%%%%%%%%%%%%%%%%%%%%%%%%

As evident from the figure, in the kinematic domain probed by SPS heavy-ion collisions, $J/\psi$ production gets enhanced due to increasing gluon densities whereas the shadowing effects in Drell-Yan production can be attributed to the depleted quark densities inside the Pb nucleus. Effects are maximum for central collisions, as expected. We have also plotted the ratio $R^{AA}_{J/\psi/DY}$ defined as the ratio of the two production cross sections in di-muon channel, in absence of any final state nuclear dissociation ($R^{AA}_{J/\psi/DY}=B_{\mu\mu}\sigma^{AA}_{J/\psi}(\varepsilon^2=0)/\sigma^{AA}_{DY}$). Due to absence of final state dissociation the ratio exhibits a decreasing trend with decrease in collision centrality. Note for $b=0$, the ratio assumes a value $R^{AA}_{J/\psi/DY} \simeq 53$. It is close to the normalization factor used in \cite{bl96} to describe the  $J/\psi/DY$ ratio in 158 A GeV Pb+Pb collisions. It might also be interesting to look at the rapidity dependence of the shadowing function. In Fig.~\ref{fig3a}, we have shown the rapidity dependence of the shadowing function for $J/\psi$ production. The contributions arising from the gluon-gluon ($gg$) fusion and quark-antiquark ($q\bar{q}$) annihilation are plotted separately along with their sum. The shadowing effects for quark annihilation at forward rapidity, is more than compensated by gluon fusion.

The reader may note that all our present calculations are based on the default EPS09 nPDF sets. With this parametrization, we are able to describe the entire Pb+Pb data within the QVZ approach. However there are many other parameterizations of nuclear parton densities, for which, in particular the gluon distributions which can not be probed directly in experiments, are of very different nature. One can perform a systematic study with different nPDF sets as inputs to check whether all of them can reproduce the Pb+Pb data equally well by compensating the difference in shadowing effects by adjusting the model parameter $\epsilon^2$ that accounts for the final state nuclear dissociation. Earlier we have seen that the $p+A$ data at different SPS energies can also be described within QVZ model using free parton pdfs~\cite{partha1}. In that case one gets relatively smaller values of $\epsilon^2$ (provided all other model parameters are fixed from $p+p$ data) due to absence of anti-shadowing effects. In \cite{Lourenco}, the authors have analyzed the $J/\psi$ production cross sections in $p+A$ collisions at different fixed target experiments, within the Glauber model framework. They used several sets of nuclear parton distributions having large difference in gluon densities for a given $x$. All of them are found to reproduce the data with different values of final state $J/\psi$ absorption cross section ($\sigma_{abs}^{J/\psi}$). Stronger is the anti-shadowing enhancement in the initial state gluon pdf, larger is the extracted $\sigma_{abs}^{J/\psi}$.  Similar effects might be seen in case of heavy-ion data as well, but one certainly needs a quantitative check to make any robust conclusion and we plan to do so in a future communication. In this context, we would also like to point out that in the original proposition of QVZ model nuclear effects on parton densities were not taken into account. Even then the model was found to describe the then available NA50 data on $J/\psi$ suppression in Pb+Pb collisions. In that case the value of $\epsilon^2$ was tuned from the heavy-ion data itself. Unfortunately a direct comparison with our $\epsilon^2$ value is not possible due to difference in the other model parameters.

%%%%%%%%%%%%%%%%%%%%%%%%%%%%%%%%%%%%%%%%%%%%%%%%%%%%%%%%%%%%%%%%%%%%%%%%%%%%%%%%%%%%%%%%%%%%%%%%%%%%%%%%%%%%%%%%%
\begin{figure*}
\scalebox{1}
{
\includegraphics[width=0.4\textwidth]{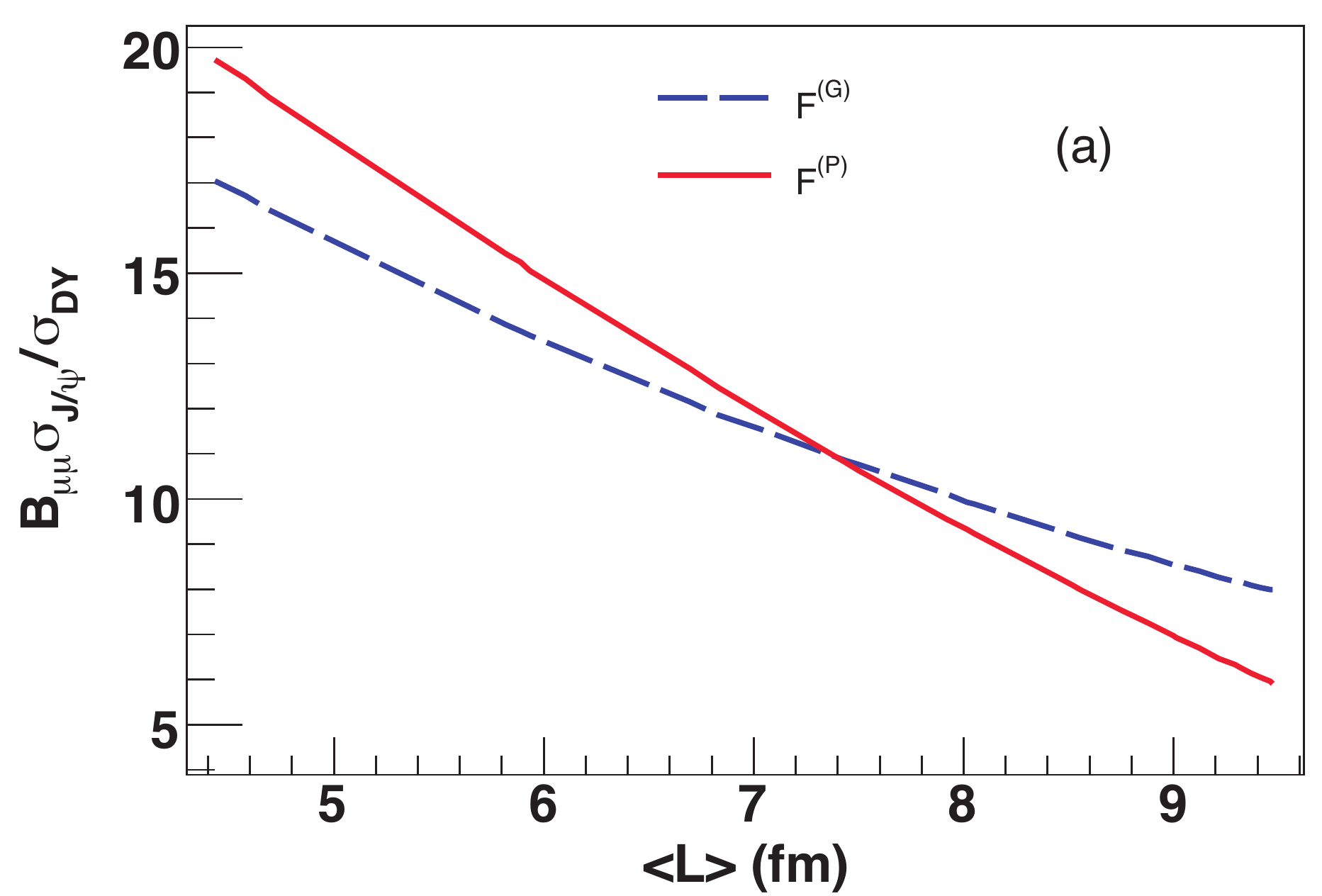}
\hspace{20pt}
\includegraphics[width=0.4\textwidth]{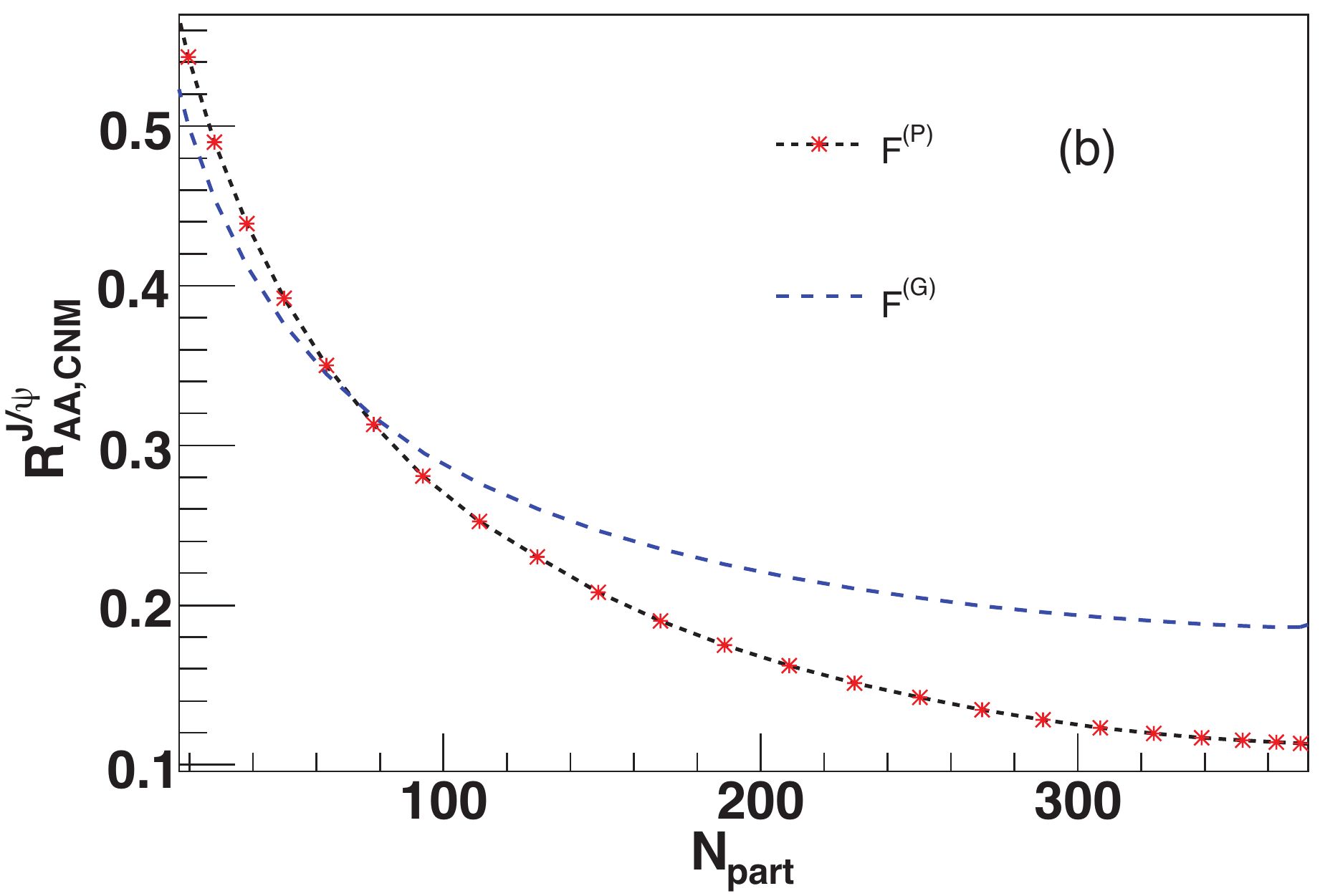}
}
\caption{\footnotesize (Color online) Model prediction for centrality dependence of the (a) $J/\psi$-to-Drell-Yan ratio and (b) nuclear modification factor of $J/\psi$ ($R_{AA}^{J/\psi}$), in 25 A GeV Au+Au collisions, at mid-rapidity. The CNM effects incorporated in the model calculations include the density dependent shadowing corrections in the initial state and final state dissociation of the pre-resonant $c\bar{c}$ pairs inside the nuclear medium. No hot medium effect is taken into account. Two theory curves, as usual correspond to the two different parametric forms of the $J/\psi$ transition probability representing different color neutralization mechanisms.}
\label{fig6}
\end{figure*}
%%%%%%%%%%%%%%%%%%%%%%%%%%%%%%%%%%%%%%%%%%%%%%%%%%%%%%%%%%%%%%%%%%%%%%%%%%%%%%%%%%%%%%%%%%%%
\section{Prediction at FAIR}

%%%%%%%%%%%%%%%%%%%%%%%%%%%%%%%%%%%%%%%%%%%%%%%%%%%%%%%%%%%%%%%%%%%%%%%%%%%%%%%%%%%%%%%%%%%%

In this section we will extrapolate our model to make predictions on parton shadowing and resulting centrality dependence of $J/\psi$-to-Drell-Yan ratio in nuclear collisions at FAIR. As a baseline we choose the Au+Au collisions at a beam energy 25 A GeV, that will be within the reach of the FAIR accelerators. Such extrapolations are of course based on the hope that perturbative calculations are still valid at FAIR energy domain. However in such low energy collisions quarkonium production would occur close to the threshold and factorization of the initial state into parton distribution functions may not be valid (see for example \cite{partha3} for an alternative approach to calculate near-threshold quarkonium production based on scaling theory). But the test of the QCD factorization in the near threshold production can only be performed once we have data. Let us first investigate the behavior of the shadowing functions for $J/\Psi$ and Drell-Yan productions in the FAIR energy domain. Inclusive cross sections for $J/\psi$ production is obtained by integrating the single differential distribution over the rapidity range: $-0.5 \le y_{c.m.} \le 0.5$. For Drell-Yan process, double differential cross section is integrated over the mass range $2.9 \le m_{\mu\mu} \le 4.5$ GeV/$c^2$ and rapidity range: $-0.5 \le y_{c.m.} \le 0.5$.  As illustrated in Fig.~\ref{fig4}, both the $J/\psi$ as well as Drell-Yan productions exhibit shadowing, with Drell-Yan production showing stronger effects, where the production cross sections in central Pb+Pb collisions reduces by almost $50 \%$ compared to $p+p$ collisions. The way these shadowing and anti-shadowing corrections affect the $J/\psi$ to Drell-Yan ratio at different collision energies, can be understood better if one looks at the evolution of the nuclear parton densities. Fig.~\ref{fig5} shows the variation of EPS09 nuclear PDFs inside Au nucleus, evaluated at a momentum scale $Q=3.1$ GeV, suitable for $J/\psi$ production, as a function of momentum fraction, $x$. Since at SPS and FAIR energies production is dominantly at  mid-rapidity, so let us consider the region $y_{c.m.} \approx 0$, which would correspond to the domain $x_1 = x_2 =x = Q/\sqrt{s_{NN}}$. Hence at SPS ($\sqrt{s_{NN}} \approx 17.3$ GeV), we have $x_{SPS} \approx 0.18$, where both the gluons and valence quarks densities exhibit anti-shadowing whereas the sea quarks show shadowing effects. This is finally reflected in $S_{J/\psi} > 1$ and  $S_{DY} <1$ at SPS energies. On the other hand, the upcoming expriment at FAIR energies ($\sqrt{s_{NN}} \approx 6.9$ GeV), would probe the kinematic region $x_{FAIR} \approx 0.45$. In this regime,  gluon, valence quark as well as sea quark densities inside the nuclei get depleted (the EMC effect). Due to higher depletion of sea quarks, Drell-Yan process shows stronger reductions. Also even in the absence of the final state dissociation of the pre-resonance $c\bar{c}$ pairs, the initial state shadowing effects alone lead to a 10 -15 $\%$ suppression $J/\psi$ production in Pb+Pb collisions. Eventually this also leads to lower values for the ratio $R^{AA}_{J/\psi/DY}$ compared to the SPS. 

We would like to end this section with quantitative  predictions for $J/\psi$-to-Drell-Yan ratio in presence of CNM suppressions, which can be directly contrasted with the data when available. In case the data for Drell-Yan production suffers from lack of sizable statistics, the suppressions can also be characterized in terms $R_{AA}$. Most of the present day experiments like those at RHIC or LHC are publishing their results on charmonium suppression in terms of $R_{AA}$. At FAIR energies also one can express the data in terms of $R_{AA}$, though our wish in that case would be a direct measurement of charmonium production cross sections in $p+p$ collisions, rather than the extrapolation of the $p+A$ results. In Fig.~\ref{fig6} we thus present our model predictions for centrality dependence $J/\psi$ suppression in Au+Au collisions both in terms of $J/\psi$/DY ratio as well as $R_{AA}$. The two theory curves, as usual correspond to two different parameterizations of $F(q^2)$. The corresponding values of $\epsilon^{2}$ accounting for final state dissociation are obtained from the beam energy dependence of $\epsilon^{2}(E_b)$, parametrized in~\cite{partha1}. Note that at SPS energy domain, the entire data corpus for 158 A GeV Pb+Pb collisions, can be described by CNM effects alone, provided we opt for power-law ($F^{(P)} (q^2)$) form for $J/\psi$ transition probability. If the data those will be collected at FAIR would show smaller values of the ratio compared to the present estimations, then such additional suppressions can be attributed to the presence of a secondary medium, characterization of which will be of further theoretical interest. 

\section{Summary}
In summary, we have discussed the $J/\psi$-to-Drell-Yan ratio in $p+A$ and $A+A$ collisions, measured by NA50 collaboration at SPS, indicating a significant role that the nuclear modification of the parton densities may have on these ratios. The $J/\psi$ cross sections are mainly sensitive to the gluon distributions, whereas the Drell-Yan cross sections are dominated by $q\bar{q}$ annihilation. As the nuclear effects in parton distributions are in general different for quarks and gluons, they are found to induce a relatively large effect on $J/\psi$-to-Drell-Yan ratio. The nuclear dissociation of the pre-resonant $c\bar{c}$ pairs is also taken into account by our calculations. Once the nuclear-modified parton densities inside the target and projectile nuclei are correctly taken into account,our analysis gives a reasonable description of data for both $p+A$ as well as Pb+Pb collisions. Thus it leaves no room for any additional suppression mechanism to set in even for most central Pb+Pb collisions, where formation of a hot and dense secondary medium is highly envisaged. One might note that although the pattern describing the centrality dependence of $\sigma_{J/\psi}/\sigma_{DY}$ remains essentially unaffected, the numerical values of the ratio are somewhat sensitive to the specific choice of the pdf. This happens because different pdfs give slightly different shape of DY di-muon mass distribution and consequently lead to a different DY yield in the region $2.9 \le m_{\mu\mu} \le 4.5 $ GeV/c$^2$. For consistency one should use a single set of pdf for all data samples. In our calculations, we have used LO central MSTW set for free proton pdfs and LO EPS09 default interface for the nuclear effects. The errors in the pdf sets are ignored in the present calculations. A more detailed calculations should take such uncertainties into account though they are not expected to produce large effects so to change our main results. MSTW takes into account the iso-spin asymmetry of the quark sea ($\bar{u} \ne \bar{d}$). In case of nuclear collisions, the shadowing factors are assumed to be proportional to the local nuclear density which gives rise to the new source of impact parameter dependence of the $J/\psi$-to-Drell-Yan ratio. Explicit evaluation of the shadowing factors in Pb+Pb collisions, show that in the kinematic domain probed by SPS, $J/\psi$ production is anti-shadowed whereas Drell-Yan production undergoes shadowing. Model is extrapolated to predict the centrality dependence of $J/\psi$-to-Drell-Yan ratio in Au+Au collisions at FAIR. In this regime both $J/\psi$ and Drell-Yan productions are found to undergo suppression due to depletion of nuclear parton densities. In order to determine the final state effects in charmonium production, such initial state modifications must evidently be brought under control. Data from the future FAIR accelerators with unprecedented high beam intensities are thus highly welcome. One can be hopeful to minimize statistical errors prevalent for Drell-Yan pairs, at FAIR, provided the data will be collected for a fairly long period. A precise determination of the nuclear effects will also help to distinguish the possible effects of a compressed baryonic matter anticipated at FAIR energy collisions. 

\section{Acknowledgement}
It is a pleasure to thank Prithwish Tribedy for his kind help during the preparation of the manuscript. PPB would also like to acknowledge the services provided by the grid computing facility at VECC-Kolkata, India for facilitating to perform a major part of the computation used in this work

\end{document}